\documentclass[]{article}
\usepackage{emulateapj,onecolfloat,graphicx}

\def\jcoph{J. Comp.\ Phys.}

\def\cf{{\it cf.}}
\def\eg{{\it e.g.}}
\def\etal{{\it et al.}}

\def\ie{{\it i.e.}}
\def\nb{{\it n.b.}}

\def\DF{{\sc DF}}
\def\Dv2{\hbox{$\Delta_{v/2}$}}
\def\Mb{M_b}

\def\bL{\;\pmb{\mit L}}
\def\pmb#1{\setbox0=\hbox{$#1$}%
  \kern-0.25em\copy0\kern-\wd0
  \kern.05em\copy0\kern-\wd0
  \kern-0.025em\raise.0433em\box0}
\long\def\Ignore#1{\relax}

\slugcomment{\it Rutgers Astrophysics Preprint \#453, To appear in ApJ}

\begin{document}

\twocolumn[
\title{Bar-Halo Friction in Galaxies III: Halo Density Changes}
\author{J. A. Sellwood}
\affil{Rutgers University, Department of Physics \& Astronomy, \\
       136 Frelinghuysen Road, Piscataway, NJ 08854-8019 \\
       {\it sellwood@physics.rutgers.edu}}

\begin{abstract}
The predicted central densities of dark matter halos in LCDM models
exceed those observed in some galaxies.  Weinberg \& Katz argue that
angular momentum transfer from a rotating bar in the baryonic disk can
lower the halo density, but they also contend that $N$-body
simulations of this process will not reveal the true continuum result
unless many more than the usual numbers of particles are employed.
Adopting their simplified model of a rotating rigid bar in a live
halo, I have been unable to find any evidence to support their
contention.  I find that both the angular momentum transferred and the
halo density change are independent of the number of particles over
the range usually employed up to that advocated by these authors.  I
show that my results do not depend on any numerical parameters, and
that field methods perform equally with grid methods.  I also identify
the reasons that the required particle number suggested by Weinberg \&
Katz is excessive.  I further show that when countervailing
compression by baryonic settling is ignored, moderate bars can reduce
the mean density of the inner halo by 20\% -- 30\%.  Long, massive,
skinny bars can reduce the mean inner density by a factor $\sim 10$.
The largest density reductions are achieved at the expense of removing
most of the angular momentum likely to reside in the baryonic
component.  Compression of the halo by baryonic settling must reduce,
and may even overwhelm, the density reduction achievable by bar
friction.
\end{abstract}

]

\section{Introduction}
\label{intro}
The LCDM model for the formation of structure and galaxies in the
universe makes specific predictions about the density profiles of
galaxy halos.  It is generally reported that the spherically averaged
density profile approximates a broken power law of the form
\begin{equation}
\rho(r) = {\rho_s r_s^3 \over r^\alpha(r+r_s)^{3-\alpha}},
\label{NFW}
\end{equation}
with $\rho_s$ and $r_s$ setting the density and spatial scales, and $1
\la \alpha \la 1.5$.  The NFW profile (Navarro, Frenk \& White 1997)
has $\alpha=1$, but recent work supports larger values (\eg, Diemand
\etal\ 2004).  Power \etal\ (2003) and Navarro \etal\ (2004) suggest
that the inner profile slope decreases continuously towards smaller
radii, but the logarithmic slope remains $\la -1$.

The halo concentration is defined as $c = r_{\rm out}/r_s$, with the
outer radius, $r_{\rm out}$, being that inside of which the mean
density, in units of the cosmic closure density, is $\bar\delta_{\rm
out}$; commonly $\bar\delta_{\rm out} = 200$.  The concentration, $c$,
can readily be related to $\rho_s$ by integrating eq.~(\ref{NFW}).
Its mean value, which varies slowly with halo mass, is a second major
prediction of the simulations (\eg, Bullock \etal\ 2001, but
see also Neto \etal\ 2007).

Attempts to estimate the dark matter density profiles in galaxies
directly are beset by many observational and modeling difficulties
(\eg, Swaters \etal\ 2003; Rhee \etal\ 2004; Valenzuela \etal\ 2007).
Alam, Bullock \& Weinberg (2002) therefore proposed a quantity that is
less sensitive to observational uncertainty, though still based on the
spherically averaged mass distribution.  They define \Dv2\ to
be the mean halo density, normalized by the cosmic closure density,
interior to the radius at which the circular speed of the halo alone
rises to half its maximum value.  As this radius is typically a few
kiloparsecs from the center of a galaxy, the quantity is less
sensitive to observational, or numerical, uncertainties.  The quantity
is easily extracted from simulations, and can be estimated from
high-quality observational data, if the baryonic contribution to the
central attraction is known, or can be neglected.

\begin{figure}[t]
\includegraphics[width=\hsize]{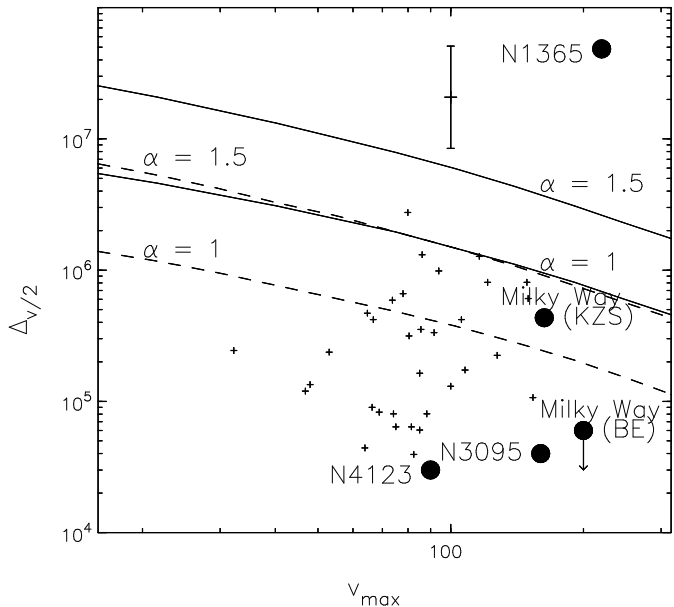}
\caption{\footnotesize After Alam \etal\ (2002).  Solid and dashed
lines show \Dv2\ predicted from two different parameter sets
for LCDM and for two different values of the slope of the inner
density profile.  The error bar indicates the approximate spread in
these predicted values.  Plus symbols show the galaxy data collated by
Alam \etal, and the sources for the five large labeled points are
described in the text.}
\label{Weiner}
\end{figure}

A major advantage of \Dv2\ is that it does not require any
assumption to be made about the halo density profile.  However, it may
be useful to note that for the NFW halo, $\alpha = 1$ in
eq.~(\ref{NFW}), we have $r_{v/2} \simeq 0.127r_s$, and $\Dv2
= 3.36 \bar\delta_{\rm out} c^3 /[\ln(1+c) - c/(1+c)]$.

I have redrawn the principal figure of Alam \etal\ as
Figure~\ref{Weiner}.  The plus symbols show the points collated by
those authors from fits to galaxies for which the baryonic
contribution was assumed to be negligible.  The points for NGC~4123
and NGC~3095 are from Weiner (2004), while that for NGC~1356 is from
Z\'anmar S\'anchez \etal\ (2008) using the same method.  I plot two
points for the Milky Way: the upper point is model B$_1$ from Klypin,
Zhao \& Somerville (2002), while the lower shows the upper limit from
Binney \& Evans (2001) that the maximum halo contribution at the solar
circle ($r=8\;$kpc) is $100\hbox{ km s}^{-1}$.  For this latter model,
I adopted $v_{\rm max} = 200\hbox{ km s}^{-1}$ in the Milky Way for
the abscissa, but the ordinate does not depend on this assumption,
since Binney \& Evans argue that the halo density cannot increase
steeply towards the center.  It is unclear how these two separate
models could be reconciled.

Predictions from two separate LCDM models are also reproduced from
Alam \etal \ The solid lines in Fig.~\ref{Weiner} show the predicted
values of \Dv2\ when $\Omega_m=0.3$, $h=0.7$, $\sigma_8=1$,
$n=1$, and for values of $\alpha=1$ \& 1.5.  The error bar indicates
their estimated factor $\sim 2.5$ spread in the predicted values of
\Dv2.  The recent WMAP results (Spergel \etal\ 2007) require
a lower $\sigma_8$ and also suggest that the initial power spectrum of
density fluctuations is not scale free, as assumed for the solid
lines, but may be tilted with less power on small scales.  Zentner \&
Bullock (2002) have already shown that power spectra of this form lead
to halos of lower concentration, and the predictions for one such
model ($\Omega_m=0.4$, $h=0.65$, $\sigma_8=0.7$ and $n=0.93$) adopted
by Alam \etal\ are shown by the dashed lines.  Modern data (\eg,
Tegmark \etal\ 2006) indicate a slightly higher $\sigma_8$, suggesting
that the dashed lines are on the low side.

The data points in this plot are not in good agreement with the
predictions, especially since simulations suggest $\alpha>1$.  Note
that three of the large points, which are based on detailed models for
each galaxy, are among the most discrepant, and that the discrepancy
for these baryon-dominated galaxies will widen by at least a factor of
a few when halo compression by baryonic infall is taken into account.
The particular tilted spectrum model shown by the dashed lines reduces
the discrepancy between the prediction and the data, but does not
eliminate it.  Dynamical friction constraints from Debattista \&
Sellwood (2000) lend support for low dark matter densities in barred
galaxies.

The point for NGC 1365 is from an NFW halo with concentration
$c_{200}=61$, which Z\'anmar S\'anchez \etal\ (2008) determined to
yield the best-fit to their data.  I plot this point from the current
compressed halo in order to be consistent with the other points that
also show the compressed halos.  Z\'anmar S\'anchez \etal\ estimate
that an initial halo with $c_{200}=22$ would yield an acceptable fit
after compression, although the baryon fraction in this case is a very
high 27\%.  This uncompressed halo is much closer to the LCDM
predictions, and has $\Dv2 = 3.5 \times 10^6$, but most discrepancies
would worsen were halo compression taken into account for all other
galaxies also.

Low central densities of DM in galaxies today need not be a problem
for LCDM if the cusps can be erased subsequently during galaxy
formation or evolution.  Several ideas to reduce the central DM
density have been proposed:

\begin{itemize}
\item Binney, Gerhard \& Silk (2001), and others have proposed that
the halo profile is altered by adiabatic compression as the gas cools
followed by impulsive outflow of a large fraction of the baryon mass.
One possible mechanism to produce such an outflow might be supernovae
and stellar winds resulting from a burst of star formation.  The idea
was examined by Navarro, Eke \& Frenk (1997) and by Gnedin \& Zhao
(2002), who found that only a mild reduction in the central DM density
could be achieved in this way.  Gnedin \& Zhao tested the extreme case
that 100\% of the baryonic component was somehow blasted out
instantaneously, yet found that even with this deliberately extreme
assumption, the central density decreased by little more than a factor
of two, unless the initial baryons were unrealistically concentrated
to the halo center.

\item El-Zant, Shlosman \& Hoffman (2001) propose that the cusp in the
halo density can be erased by dynamical friction with orbiting mass
clumps.  In essence, this is a process of mass segregation, in which
heavy ``gas'' particles lose energy and settle to the center due to
interactions with the light DM particles.  However, Jardel \& Sellwood
(2008) show that the settling time is uninterestingly slow unless the
baryonic clumps are extremely massive.

\item Milosavljevi\'c \etal\ (2002) point out that a binary
supermassive black hole (BH) pair created from the merger of two
smaller galaxies will eject DM (and stars) from the center of the
merger remnant.  They also argue that the DM mass removed for a given
final BH mass is greater if the final BH is built up in a series of
mergers each having correspondingly lower mass BHs.  While this
mechanism must operate wherever binary BHs have been formed, the
radial extent over which the mass is reduced is rather limited.  They
predict that the cores in the DM halos could possibly be larger than
those in the bulge stars, whereas the discrepancy shown in
Fig.~\ref{Weiner} applies to much larger radii.  Furthermore, shallow
density gradients are observed in DM-dominated galaxies with
insignificant bulges (Simon \etal\ 2005; Kuzio de Naray \etal\ 2006)
which are likely to have very low-mass BHs (Gebhardt \etal\ 2000;
Ferrarese \& Merritt 2000), if they contain BHs at all.

\item Weinberg \& Katz (2002) suggest that a bar in the disk could
flatten the cusp also through dynamical friction.  Here I study this
possibility in more detail.

\end{itemize}

Bar-driven halo density changes in fully self-consistent simulations
reported so far have been minor, and of both signs.  Debattista \&
Sellwood (2000) showed a modest halo density reduction in their
Fig.~2, and Athanassoula's (2003) simulations also indicate a small
halo density decrease.  On the other hand, Sellwood (2003) and Col\'\i
n, Valenzuela \& Klypin (2006) report the opposite behavior in
simulations with more extensive halos, finding instead that loss of
angular momentum from the disk caused the halo to contract, with the
deeper disk potential well compressing the halo still more.
Holley-Bockelmann, Weinberg \& Katz (2005) however, report that the
inner cusp was flattened in most of their experiments.  While the
radial extent of the effect was modest, the cusp was erased to a
radius less than one fifth the bar semi-major axis, they continue to
insist that the effect can be important.  They further argue that the
absence of significant density reductions in some published cases is
due to numerical inadequacies.

Thus two separate issues need to be clarified.  First, what are the
numerical requirements to obtain reliable results from simulations?
and second, what physical properties of the bar affect the extent to
which the halo density can be reduced?

Weinberg \& Katz (2007a \& b; hereafter WK07a and WK07b) claim that
simulated halos should contain between tens of millions and billions
of particles to obtain the correct result.  They reach this conclusion
from perturbative calculations of the interaction between a rotating
quadrupole potential and orbits in a spherical halo.  Previous theory
(Tremaine \& Weinberg 1984) had shown that the important exchanges
occur at resonances, and while an individual halo orbit may either
gain or lose angular momentum, a net torque arises because there is a
slight excess of gainers over losers.  WK07a argue that it is
important to have an adequate density of particles in phase space in
order to obtain the correct balance, a criterion they dub
``coverage.''  They also argue that density fluctuations due to a
finite number of particles cause the orbits of particles in
simulations to deviate from those in a smooth potential, and that
particles will therefore diffuse into and out of resonances due to
such effects.  If the diffusion rate is high, the simulation will not
capture the appropriate smooth behavior, affecting the torque between
the bar and the halo particles.  They further argue that the lumpiness
of the potential due to particle fluctuations depends on the method
for calculating the gravitational field, and that field methods that
employ an expansion in a set of basis functions will be intrinsically
smoother than all other methods, and will therefore yield more
reliable results.

Studies of bar-halo interactions, the slowing of bars, and the
evolution of halo mass profiles cannot be pursued with confidence
until the issues raised by WK07a are addressed.  It is important to
check whether results from previous and future studies with the usual
${\cal O}(10^6)$ particles are, or are not, compromised by numerical
inadequacies.

In Paper I (Sellwood 2006), I demonstrated explicitly that resonant
exchanges between halo particles and the quadrupole field of a mild
bar were taking place.  I also showed that simulations both with and
without self-gravity could converge to a frictional drag that was
independent of the number of particles for feasible particle numbers.
The mild bars used in that study did not, however, cause any
significant change to the halo mass profile and did not, therefore,
represent a direct challenge to the claims by WK07a.  Other studies
(Athanassoula 2002; Ceverino \& Klypin 2007) have demonstrated the
existence of many orbits trapped in various resonances, suggesting
that particle noise does not preclude trapping from occurring, even
when $N \sim 10^6$.

In this paper, I first present (\S\ref{fiducial}) a further study of
bar-halo interactions with much stronger bars that do cause large
density reductions in order to provide a direct test of the issues
raised by WK07a.  Again I find (\S\ref{numtests}) that numerical
results are quite insensitive to the particle number and calculation
method.  As my results are at variance with the conclusions in WK07a
\& WK07b, I show (\S\ref{resonances}) that my simulations do indeed
reproduce a strong resonant response.  I also identify
(\S\ref{reasons}) the reasons why those authors reached incorrect
criteria for the number of particles needed.

I turn to the physically more interesting question of how strong and
large a bar is needed to cause a large density reduction in the inner
halo in \S\ref{physics}.  I show that large, massive, skinny bars can
indeed flatten the central cusp, as was already reported by Hernquist
\& Weinberg (1992), and confirmed in the rigid bar experiments of
Weinberg \& Katz (2002), Sellwood (2003), and McMillan \& Dehnen
(2005).  However, I also find that more realistic bars cause only
slight density reductions.  In \S\ref{reduction}, I show that the
possible changes in \Dv2\ in real galaxies are limited by the angular
momentum content of the baryons.

\section{Model set up}
In this section, I describe the numerical model I use throughout the
paper.  I choose a sufficiently simple model that others can easily
check my experiments.

\subsection{Halo}
For the unperturbed halo I employ the Hernquist (1990) profile
\begin{equation}
\rho_0(r) = {M r_s \over 2\pi r(r_s + r)^3}, 
\label{Hernquist}
\end{equation}
which has total mass $M$ and scale radius $r_s$.  I use the isotropic
distribution function (\DF) for this halo, which is also given by
Hernquist.  The density declines as $r^{-1}$ for $r \ll r_s$ and as
$r^{-4}$ for $r \gg r_s$.  It should be noted that this model differs
only in the outer power law slope from the Navarro, Frenk \& White (1997)
profile used by WK07b, but the important inner cusp is the same.

\begin{figure}[t]
\includegraphics[width=\hsize]{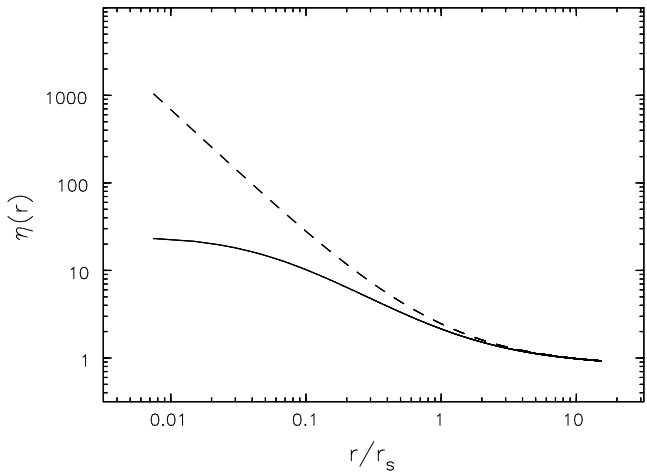}
\caption{\footnotesize The radial variation of the boost factor to the
effective number of particles when unequal particle masses are used.
The solid line is for $L_0 = 0.01$ and the dashed line for $L_0=
10^{-8}$.}
\label{nplot}
\end{figure}

While all halo particles have equal mass in most cases,\footnote{I
select particles according to the procedure described in the Appendix
of Debattista \& Sellwood (2000).} I also report experiments in which
the particles have individual masses in order to concentrate greater
numbers in the dense inner regions.  I set particle masses
proportional to a weight function $w(L) = L_0 + L$, where $L = |\bL|$
is the total specific angular momentum in units of $(GMr_s)^{1/2}$ and
$L_0$ is a constant, and select particles from the \DF\ weighted by
$w^{-1}$. Figure~\ref{nplot} plots the boost factor for the effective
number of particles $\eta(r) = {\cal N}(r)/{\cal M}(r)$, where ${\cal
N}(r)$ and ${\cal M}(r)$ are respectively the fraction of the number
of particles and the fraction of mass enclosed within radius $r$.

Choosing $L_0 = 0.01$ results in the heaviest particle being some 250
times the mass of the lightest and in half the particles being
enclosed in a sphere $r = 0.6$. A smaller sphere with $r = 0.33$
encloses the same fraction when $L_0 = 10^{-8}$, where the lightest
particle is $4 \times 10^{-9}$ of the mass of the heaviest.  As the
$N$-body codes used here are designed to simulate collisionless
dynamics, a range of particle masses should not lead to any mass
segregation.  A test, run for 100 dynamical times with no
perturbation, revealed no tendency for the small changes to either the
specific energy or specific angular momentum of the particles to
correlate with particle masses.

It is inefficient to employ many particles at large radii that take no
part in the friction process.  I therefore truncate the model by
setting the \DF\ to zero for all $E > \Phi(r_{\rm cut})$, with
$\Phi(r) = -GM/(r+r_s)$ being the gravitational potential of the
infinite Hernquist halo.  This change eliminates any particle with
sufficient energy to reach $r > r_{\rm cut}$, and the density tapers
smoothly to zero at $r = r_{\rm cut}$.  The gravitational potential
from the remaining particles is somewhat modified, and the model is no
longer an exact self-consistent equilibrium.  However, the results
presented below show that the truncation has very little effect on the
equilibrium and the density profile hardly evolves in response.  I
choose $r_{\rm cut} = 15r_s$, while the bars I employ are typically
much smaller, with semi-major axis $a \leq r_s$.  I show in
\S\ref{numtests} that the density changes in the inner halo are
unaffected by the choice of $r_{\rm cut}$ over a wide range of values.

\subsection{Bar}
In order to be able to control the bar parameters, I again employ
artificial, rigid bars (see Paper I).  The homogeneous ellipsoid has
mass $\Mb$ and axes $a:b:c$ with $a \ge b \ge c$.  It is
centered on the halo center, and rotates about its shortest axis at
angular rate $\Omega_b$.  The angular speed of the bar is adjusted to
take account of the torque from the halo, assuming it slows as a rigid
bar of moment of inertia $I=\Mb(a^2+b^2)/5$.  I use only the
(2,2) quadrupole term of the gravitational field of the bar, as
originally proposed by Hernquist \& Weinberg (1992).  I have shown in
Paper I that higher terms have a small effect, and suppression of the
monopole terms allows the bar to be introduced without affecting the
radial balance of the halo.

\begin{table}[t]
\caption{Values of $\alpha_2$ and $\beta_2$}
\label{alpbet}
\smallskip
\begin{tabular}{@{}cccc}
    $a/b$  &    $b/a$  &  $\alpha_2$  &   $\beta_2$ \\
\hline
\noalign{\smallskip}
    6.667  &      0.1500  &          17.9874  &      0.3822 \\
    6.000  &      0.1667  &          14.9372  &      0.3962 \\
    5.000  &      0.2000  &          10.6953  &      0.4225 \\
    4.545  &      0.2200  &\phantom{0}8.9196  &      0.4374 \\
    4.000  &      0.2500  &\phantom{0}6.9336  &      0.4586 \\
    3.571  &      0.2800  &\phantom{0}5.4966  &      0.4787 \\
    3.333  &      0.3000  &\phantom{0}4.7499  &      0.4917 \\
    3.226  &      0.3100  &\phantom{0}4.4255  &      0.4980 \\
    3.125  &      0.3200  &\phantom{0}4.1290  &      0.5043 \\
    3.000  &      0.3333  &\phantom{0}3.7717  &      0.5125 \\
    2.000  &      0.5000  &\phantom{0}1.3772  &      0.6059 \\
\hline
\end{tabular}
\end{table}

The approximate quadrupole field adopted by Weinberg (1985) was
designed to match that of a homogeneous bar.  I write his expression
for the bar quadrupole in spherical (not cylindrical, as mis-stated in
Sellwood 2003) polar coordinates as
\begin{equation}
\Phi_{\rm b}(r,\theta,\phi) = -{G\Mb \over a^3}
{\alpha_2 r^2 \over 1 + (r/a\beta_2)^5}\sin^2\theta e^{2i(\phi-\phi_0)},
\label{quadrup}
\end{equation}
where $a$ is the semi-major axis of the bar and $\phi_0$ is the phase
angle of the bar major axis.  I give Weinberg's prescription for
selecting the dimensionless amplitude and radius scaling parameters,
$\alpha_2$ and $\beta_2$ in the Appendix and, for fixed $a/c=10$, list
their values for the bars used here in Table~\ref{alpbet}.

I show, also in the Appendix, that this expression is a good match to
the quadrupole field of a homogeneous bar when $a/b \approx 2$, but it
gives a peak perturbation that is increasingly too strong as $a/b$ is
increased.  In Paper I, I used the exact quadrupole field, which I
added to my numerical solution for the self-consistent part of the
halo field.  As expansion of the gravitational field in multipoles on
spherical shells is not a widely used technique, such a bar field is
hard for others to reproduce.  Reproducibility therefore dictates that
I use the simple and convenient expression (\ref{quadrup}), but it
must be borne in mind that the density distribution corresponding to
this quadrupole is increasingly different from that of a homogeneous
ellipsoid having the nominal axis ratio as $a/b$ is increased.

As noted above, a fixed bar field is required in order to be able to
control the properties of the bar and address the scientific
objectives of this paper.  Bars in real galaxies do not approximate
homogeneous ellipsoids, but the quadrupole part of the field is
unlikely to differ substantially from the form (\ref{quadrup}), which
Weinberg (1985) selected in order to have the appropriate asymptotic
behavior both for $r \ll a$ and for $r \gg a$.  Few real bars have
isophotes skinnier than $a/b \sim 3$ (Reese \etal\ 2007; Marinova \&
Jogee 2007), while their mass distributions are more concentrated than
a uniform density, implying that the quadrupole field for a given
$a/b$ probably peaks interior to $r/a = (2/3)^{1/5}\beta_2$, where the
peak of the radial part of (\ref{quadrup}) occurs.  Higher multipoles
are considerably less important to the dynamics discussed here (Paper
I).  The monopole part of the bar field could be considered part of
the spherical halo, although the orbits of the particles would be
rather different.  The two most significant approximations of the
adopted bar field are that it slows as a rigid object and does not
adjust in response to the loss of angular momentum from the bar.

I introduce the bar perturbation smoothly by increasing the quadrupole
term as a cubic function of time from zero at $t=0$ to its final value
at $t=t_g$.  Tests revealed that the outcome was insensitive to the
growth-time of the bar over a broad range of values, so all
experiments reported here use $t_g = 10$ in units in which
$G=M=r_s=1$.

\subsection{Determination of the gravitational field}
In most calculations, I compute the gravitational field of the halo
particles using the radial grid method originally devised by McGlynn
(1984) with some refinements described in Sellwood (2003).  The
coefficients of a multipole expansion of the interior and exterior
masses are tabulated at a set of radii.  The default grid spacing for
these experiments places the $j$th grid shell according to the rule
$r_j = e^{\gamma j} - 1$ with $\gamma = \ln (r_{\rm max} + 1)/n$,
where $n$ is the number of radial shells and $r_{\rm max}$ is the
outer limit of the grid.  I generally use $n=300$ radial grid shells,
set $r_{\rm max} = 16r_s$, and expand up to $l_{\rm max} =4$.

This default rule for the radial grid is arbitrary, however, and I also
present results using the alternative rule $r_j = r_{\rm max}(j/n)^2$
in order to place grid points more densely in the inner parts.  In this
case, I have employed $n=1000$ radial shells.

In order to test the assertion by WK07b that field methods are
superior to all others, I present some results using the
self-consistent field (SCF) method described by Hernquist \& Ostriker
(1992), for which the fundamental function of the expansion is the
Hernquist density function (eq.~\ref{Hernquist}).  With this
procedure, I include 20 radial functions, while again expanding in
angle up to $l_{\rm max} =4$.

Expansion to low azimuthal order in both methods eliminates
small-scale variations of both the azimuthal and radial fields,
thereby hiding the graininess of the particle
distribution.\footnote{The radial grid smooths discontinuities in the
field across the radius of a source particle.} Therefore, no further
smoothing, such as gravity softening, is required for either method.

\subsection{Lop-sided instability}
I compute the motion of the halo particles in the gravitational field
arising from the particles, together with that of the external field
of the bar.  Past experience (Sellwood 2003; McMillan \& Dehnen 2005;
WK07a) has revealed that a rigid bar can drive the center of the
particle distribution away from the bar center, which results in
unphysical evolution.  Special precautions are therefore needed to
keep the particle distribution centered on the bar.  Since I compute
the field of the halo particles by a surface harmonic expansion on
spherical shells, it is simplest to eliminate only the $l=1$ terms
from the field determination, which is sufficient to ensure that the
distribution of forces is always point symmetric about the origin and
no lop-sidedness can develop.

WK07b, who employ an SCF-type method, keep the $l=1$ term active but
include the unchanging monopole term of the bar in order to inhibit
growing asymmetries in the particle distribution, as did McMillan \&
Dehnen (2005) in some of their experiments.  Not only does this
stratagem complicate the creation of the initial equilibrium, it also
introduces a rigid mass component that inhibits the collective effects
responsible for cusp flattening.  Furthermore, WK07b report that their
results are unaffected by the omission or inclusion of the $l=1$
terms; eliminating the dipole contribution to self-gravity is
therefore the simplest way to suppress this artifact.  (This stratagem
is easy with a field or grid method, but not for a tree code.
McMillan \& Dehnen describe how a tree code needs to be adapted in
order to prevent unphysical behavior when rigid bars are employed.)

\subsection{Other details}
Unless otherwise stated, the simulations reported here employ $10^6$
equal mass particles that move with a basic time step of
$0.005(r_s^3/GM)^{1/2}$, the radial grid has 300 spherical shells, and
I expand the density distribution of the particles using only the $0
\leq l \leq 4$ terms, with the $l=1$ term suppressed.  These choices
of parameters are justified in \S~\ref{numtests}.

As the orbital frequencies of particles decrease strongly with
increasing radius, I employ the multi-zone time step scheme descibed
in Sellwood (1985).  I use 5 time-step zones with the step size
increasing by a factor 2 from zone to zone; \ie\ the outermost
particles are stepped forward once for every 16 steps taken by the
innermost particles.  The contributions to the gravitational field
from slowly moving particles are interpolated in time as needed when
accelerating particles in the inner zones.

I adopt units such that $G=M=r_s=1$.

In order to estimate the halo mass profile at any time, I sort the
particles in radius and record the radius of every $n$th particle.  An
estimate of the density is the mass of the $n$ particles between these
two radii, divided by the volume of the spherical shell containing
them, and I assign this value to be the density at the mid-point of
that radial range.  I reduce the noise in this estimate by combining
multiples of $n$ particles over the bulk of the model.

\begin{figure}[t]
\includegraphics[width=\hsize]{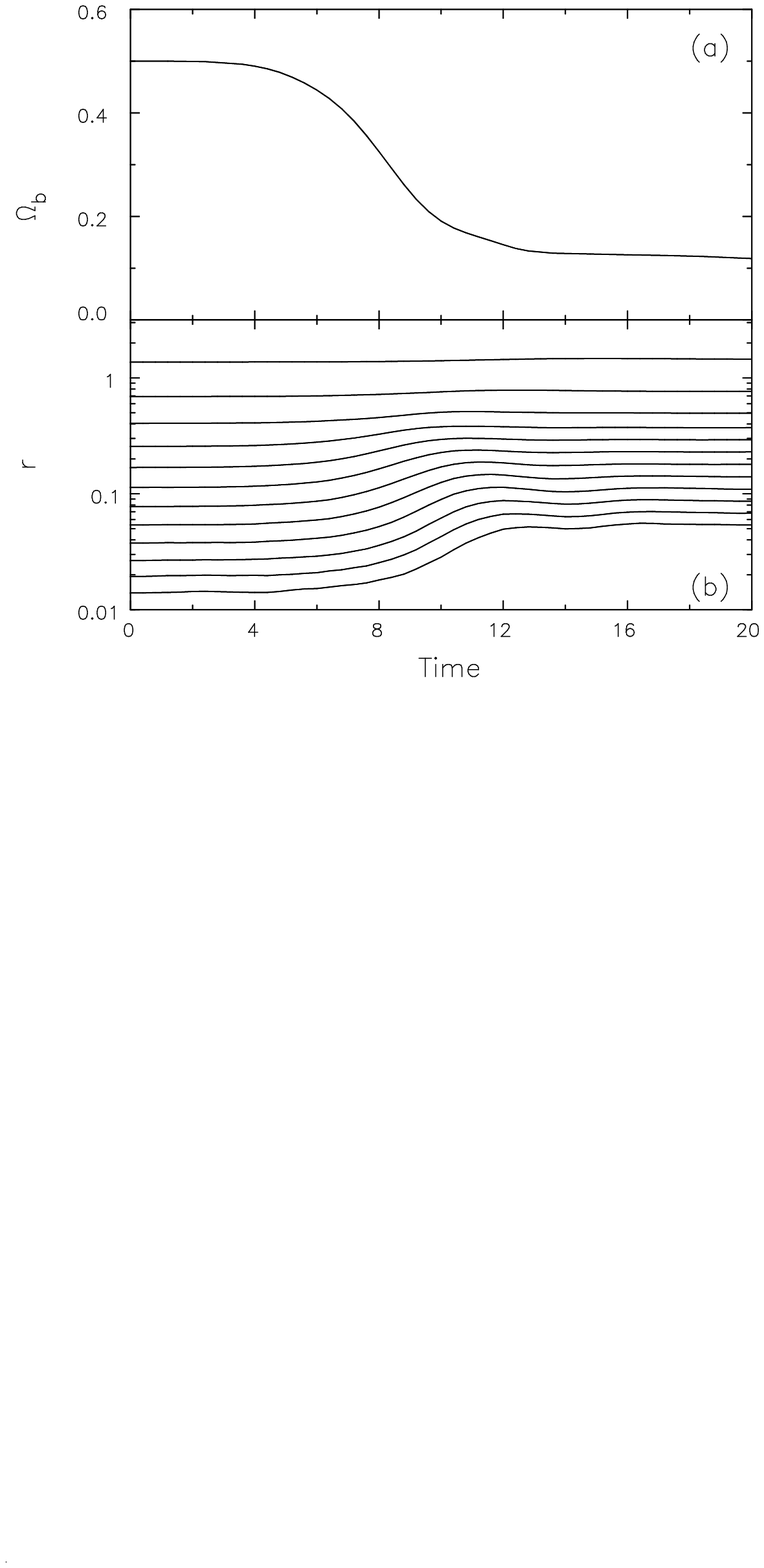}
\includegraphics[width=\hsize]{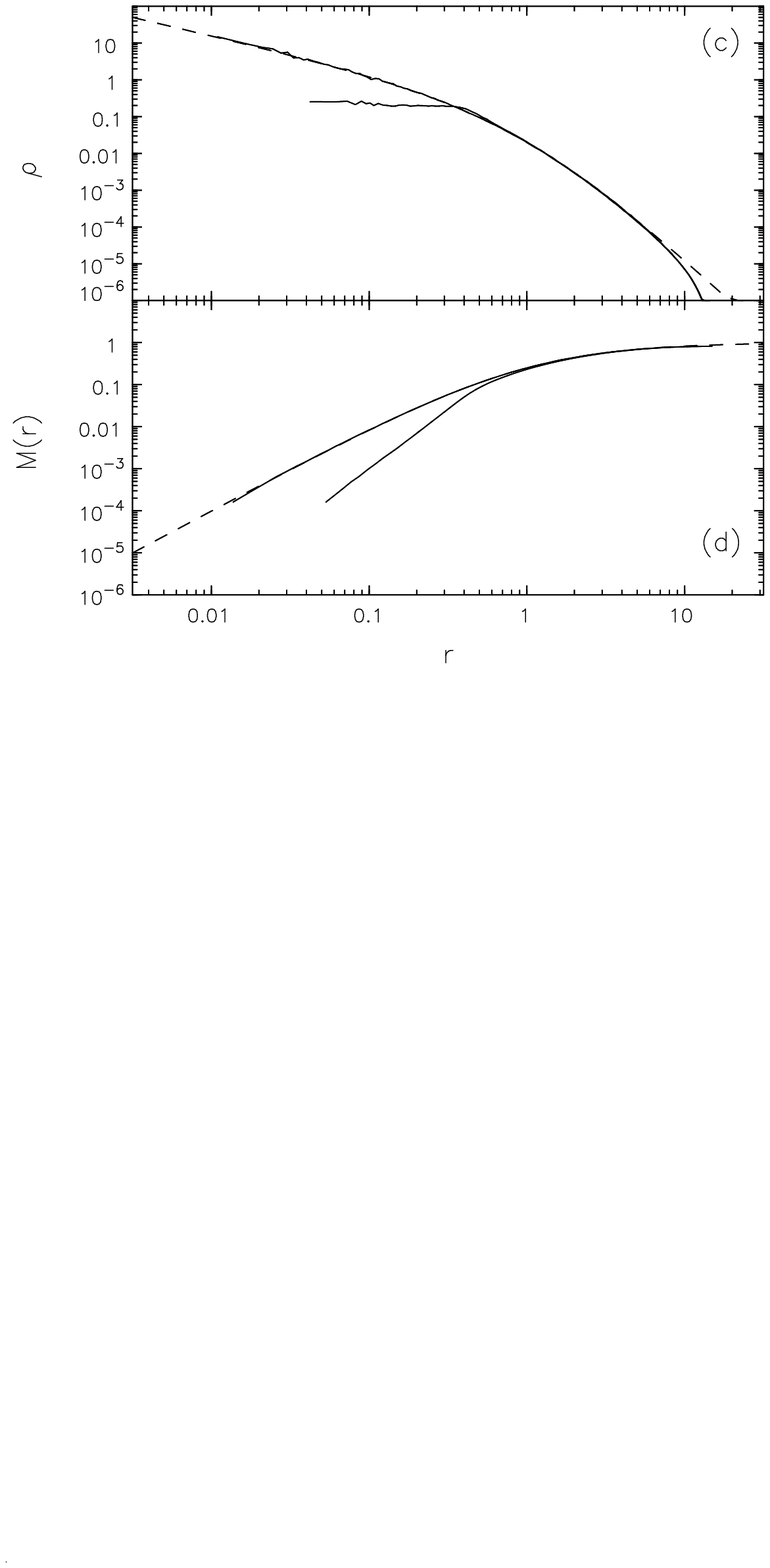}
\caption{\footnotesize The time evolution of (a) the bar pattern speed
and (b) the radii containing different mass fractions in the fiducial
run.  The smallest radius is that containing 200 particles, or
$1/5\,000$th of the mass in particles, and the mass fraction is
successively doubled for each subsequent trace.  The initial and final
density, (c), and mass, (d), profiles in the same run; the solid lines
are measured from the particles while the dashed lines show the
theoretical profile (\ref{Hernquist}).  Note that the decreased innner
density requires that the mass enclosed (d) cannot meet up with the
unperturbed mass profile until a larger radius than where the cusp in
(c) is flattened.}
\label{mrplot}
\end{figure}

\section{A Fiducial Model}
\label{fiducial}
Following WK07b, I first present a fiducial model in which the bar has
a semi-major axis $a=r_s$, a mass of half that of the halo enclosed
within $a$ so that $\Mb = 0.125M$, and the initial pattern speed is
set to place corotation at the bar end, \ie\ $\Omega_b=0.5$ with the
initial bar rotation period $=4\pi$ time units.  The nominal axis
ratio is $a:b:c = 1:0.2:0.1$, although the actual quadrupole field
employed in the simulation is stronger than that of this ellipsoid
(see Appendix).  Thus the bar is unrealistically large, massive, and
skinny, but it makes a useful starting point since WK07b correctly
argue such a model should be very easy to simulate.

The time evolution of the model is shown in Figure~\ref{mrplot}.
Friction with the halo particles, which results from resonant
interactions as described in Paper I and \S~\ref{resonances} below,
causes the pattern speed (Fig.~\ref{mrplot}a) to start to decrease as
the perturbation amplitude grows.  The bar amplitude reaches its final
value at $t=10$; the bar pattern speed is dropping very rapidly at
this time, but levels out later to about 25\% of its initial value.

The halo mass profile (Fig.~\ref{mrplot}b) does not change at first,
confirming that the model is an excellent initial equilibrium, despite
the truncation at $r_{\rm cut}$.  However, the central density
undergoes a rapid decrease over the time interval $8 \la t \la 12$,
after which further changes are comparatively minor.  Continuation of
the evolution beyond $t=20$ revealed little further change, and it is
therefore reasonable to describe the simulation at $t=20$ as
representing its final state.

Fig.~\ref{mrplot}(c) shows the initial and final density profiles.  As
estimates of density from the finite number of particles always suffer
from some noise, I plot the much more robust measure of the mass
enclosed as a function of radius in Fig.~\ref{mrplot}(d).  Initially,
$M(r) \propto r^2$ in the cusp, while the almost homogeneous core at
later times has $M(r) \propto r^3$ in the inner parts.  These curves
are measured directly from the radial distribution of particles with
no smoothing, indicating that the monopole part of the potential
derived from the particles cannot suffer from significant
fluctuations.

\begin{figure*}[p]
\includegraphics[width=.4\hsize,angle=270]{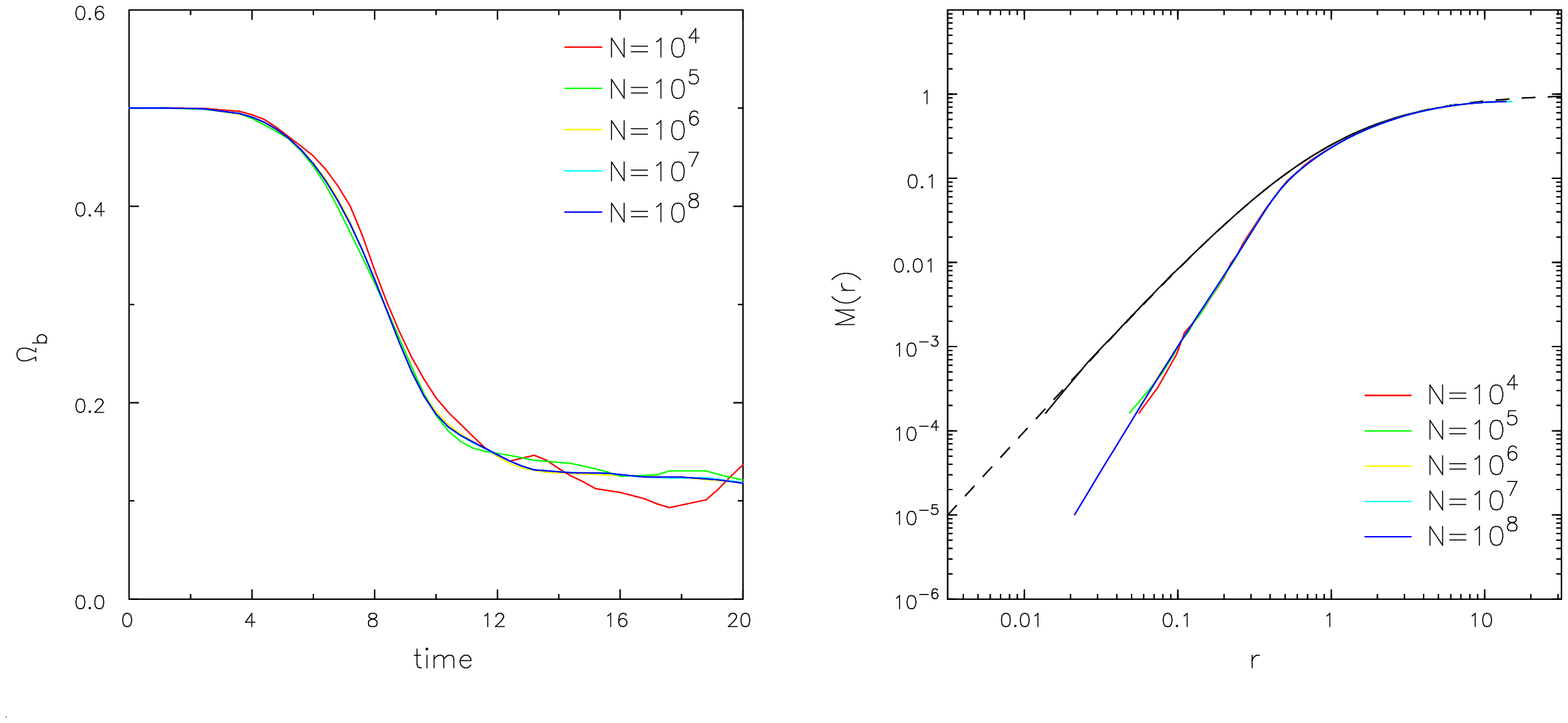} \\
\includegraphics[width=.4\hsize,angle=270]{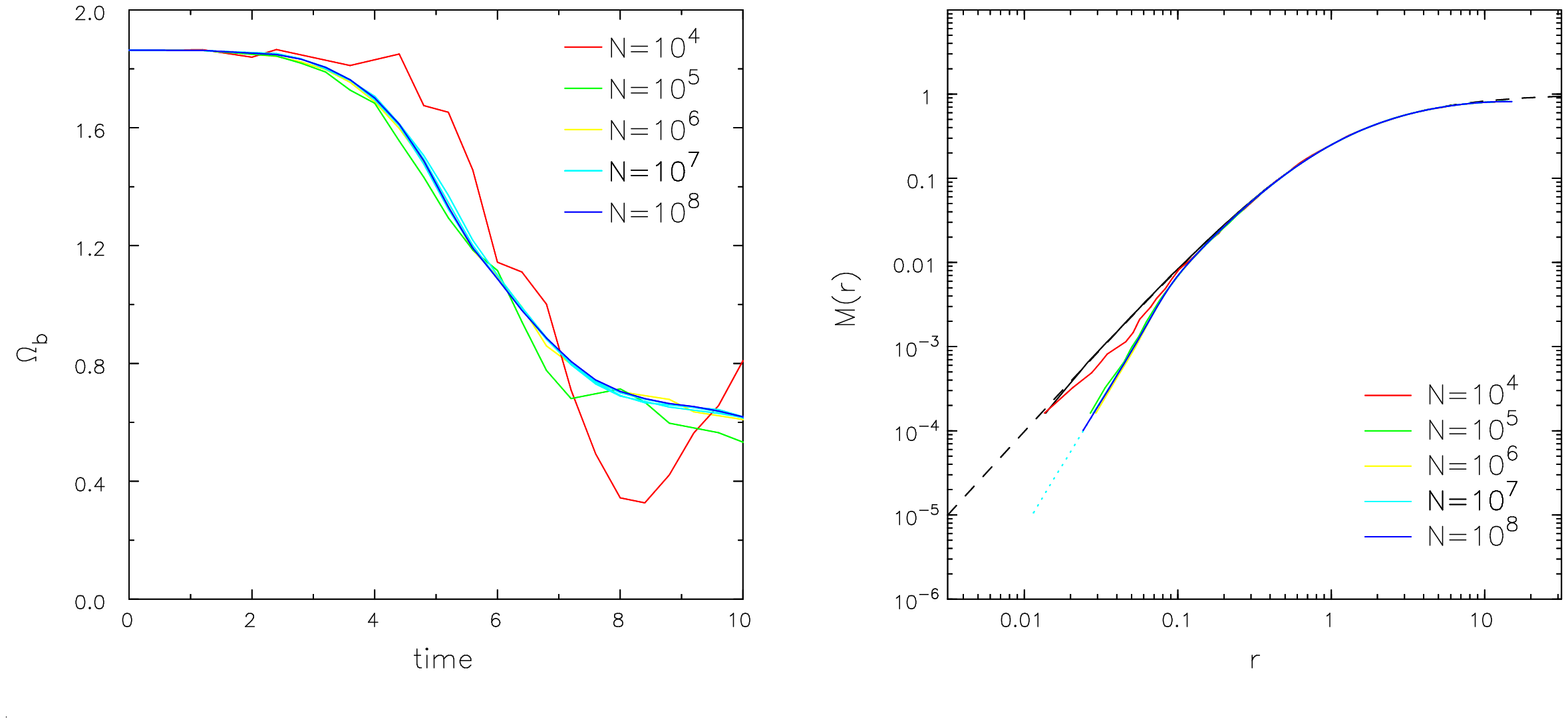} \\
\includegraphics[width=.4\hsize,angle=270]{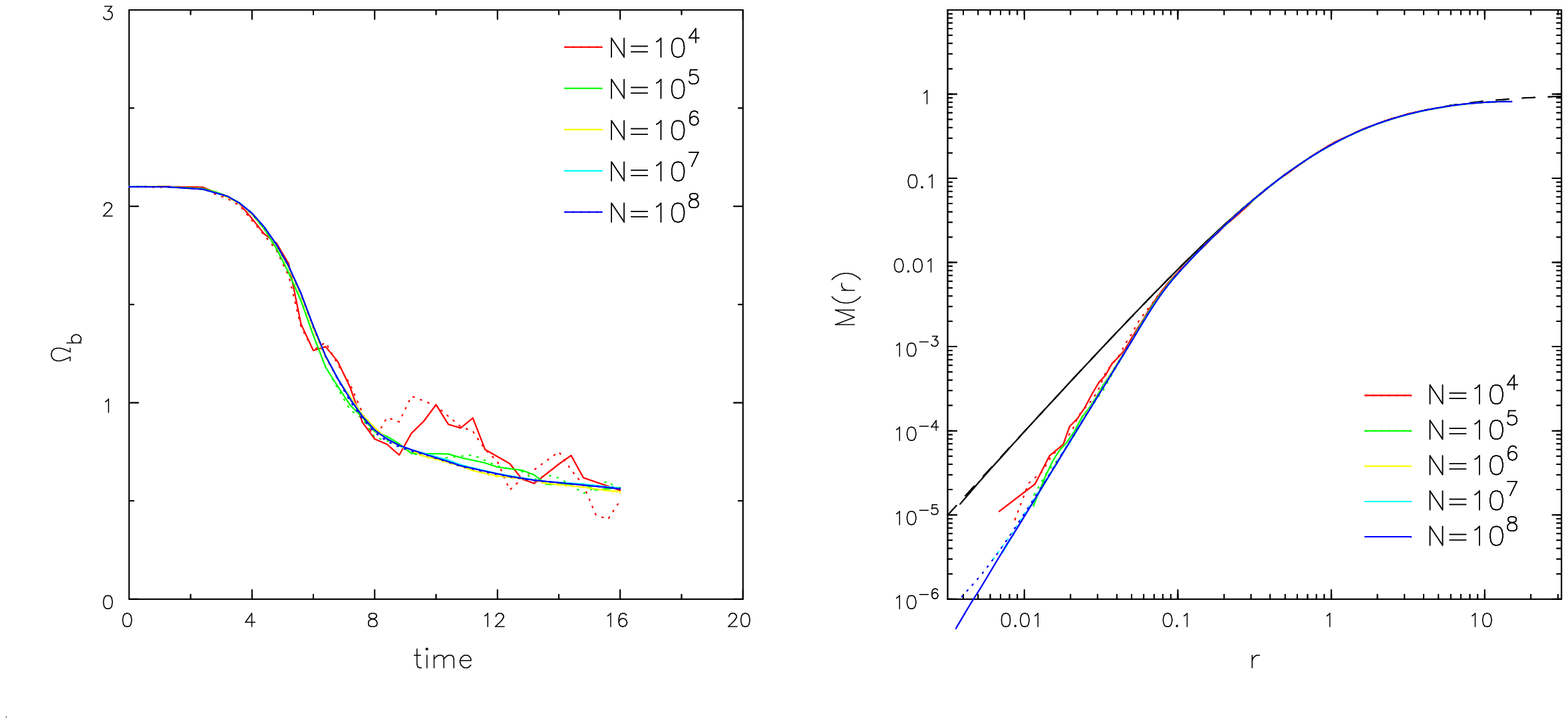}
\caption{\footnotesize The pattern speed evolution, left-hand panels,
and initial and final mass profiles, right-hand panels, in three
series of simulations in which the number of particles is varied.  The
top two rows show results using a grid method only and mostly equal
mass particles.  The bar length used in top panels is $a=r_s$ and in
the middle panels $a=r_s/5$.  The dotted curves in the middle panels
show a results with $10^7$ unequal mass particles.  The bottom panels
are all for unequal mass particles and a still shorter bar with
$a=r_s/6$; solid curves show results with a grid method, while dotted
curves were obtained using a field method.}
\label{converge}
\end{figure*}

It should be noted that the density change shown in
Fig.~\ref{mrplot}(c) is {\it larger\/} than that reported by WK07b in
a similar experiment.  As my result agrees with those found earlier
(Hernquist \& Weinberg 1992; Weinberg \& Katz 2002; Sellwood 2003;
McMillan \& Dehnen 2005) and with those from other experiments with
the NFW mass profile (not reported here), other differences in their
physical model, such as the rigid monopole term, are the likely cause.

\section{Numerical checks}
\label{numtests}
Here I present a number of checks of this and other results that are
designed to address some of the numerical concerns raised by WK07a and
WK07b.  In all cases, the bar mass is set to be half the enclosed halo
mass at $r=a$, \ie\ $\Mb = 0.5 Ma^2/(r_s+a)^2$, and the initial
pattern speed places corotation at the bar end, \ie\ $\Omega_b =
(GM/a)^{1/2}/(r_s+a)$.

\subsection{Particle number}
Figure~\ref{converge} presents results from two series of experiments
in which the number of equal-mass particles is varied over the range
$10^4 \leq N \leq 1.6 \times 10^8$ for (top row) a large bar ($a=r_s$)
and (middle row) a short bar ($a=r_s/5$).  The evolution of the bar
pattern speed and change in the mass profile are insensitive to the
particle number as long as $N \ga 10^5$; $N=10^4$ even seems adequate
for the larger bar -- the mass profile is less smooth but the
reduction in density clearly does not differ significantly.  It is
worth noting that WK07b estimate that the large bar case requires
$10^8$ equal-mass particles to obtain the appropriate behavior,
whereas my result with $N>10^8$ is no different from that with three
orders of magnitude fewer.\footnote{My model is not identical to that
employed by WK07b, but is close enough for the particle requirement to
be similar.  The unperturbed potential, the \DF\ and the dimensionless
frequencies are very similar in the cusps of both the Hernquist and
NFW halos and, if anything my bar perturbation is stronger than that
they used, which reduces the required particle number.}

The convergence in Fig.~\ref{converge} is exquisite; the different
curves show direct measurements from the simulations without
smoothing.  Yet curves for the largest $N$ mostly overlay, and
therefore obscure, those for the next largest $N$, and differences
become visible only for much smaller $N$.  WK07a correctly argue that
if the phase space coverage were inadequate, exchanges at resonances
would depend upon the few particles that happened to occupy the
resonance, making the net balance between gainers and losers
stochastic, and the resulting evolution could not converge as
impressively as shown in Fig.~\ref{converge}.  Repeated calculations
of the large bar case with different random seeds reveal some slight
stochastic behavior when $N=10^4$, but the evolution of the pattern
speed and change to the mass profile is practically identical in
another set of runs with $N=10^6$, as should be expected from the
impressive data in Fig.~\ref{converge}.

The dotted curves in the middle row are from a run with unequal mass
particles ($L_0=10^{-8}$), the alternative grid spacing rule and half
the standard time step.  The larger number of particles near the
center allows the mass profile to be traced to smaller radii, but
otherwise these refinements have no effect on either the pattern speed
or mass profile evolution.

The bottom row of Fig.~\ref{converge} is for a still shorter bar, this
time with unequal mass particles selected with $L_0=0.01$ (see \S2.1)
and with a slightly rounder bar ($a/b = 4$).  The results shown by
the solid curves were obtained using a grid method, while the dotted
curves were obtained using the SCF method.  The results from the two
methods can barely be distinguished in most cases.  It is clear that
using unequal mass particles leads to convergence at a smaller $N$ in
this numerically still more challenging case compared with that shown
in the middle row.

WK07b report results from two experiments with $a =r_s/6$ that are
similar to those in the bottom row of Fig.~\ref{converge}.  Using
unequal mass particles, they find a greater density reduction with
$N=5 \times 10^6$ than with $N=10^6$, which they attribute to the
improved numerical quality of the slightly larger $N$ experiment.  My
experiments are not an exact match to theirs; the most important
difference is their inclusion of the fixed monopole term of the bar,
but the quadrupole field of their 5:1 bar appears to be weaker than I
would employ for the same axis ratio (see Appendix), which is the
reason I used the weaker quadrupole of a 4:1 bar.  Because of these
differences, the comparison with their work is not exact, but it is
clear that I find no change in the outcome for $N \geq 10^6$ and only
a minor difference at $N=10^5$.

\subsection{Grid and field methods}
WK07b expect field methods to be intrinsically less noisy than
other techniques, yet I obtain practically identical results using
either the SCF or a grid method (Fig.~\ref{converge}, bottom row).

It should be noted that Hernquist \& Ostriker (1992) also expected
their field method to yield a slower relaxation rate than found by
other methods, but were disappointed to find that individual particle
energy variations in simulations of equilibrium spherical models
computed by the SCF method were just about as large as those for many
other methods.  Thus my finding that the evolution is independent of
the method used to calculate the forces was expected.  (See also
\S\ref{discmeth}.)

\begin{figure}[t]
\includegraphics[width=\hsize]{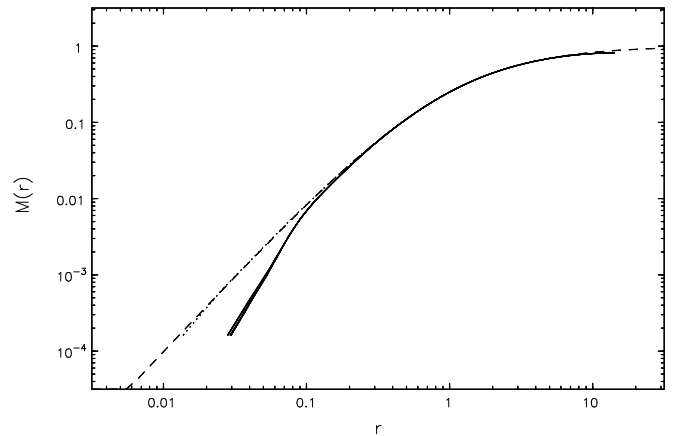}
\caption{\footnotesize The initial (dotted line) and final (solid
lines) mass profiles in a series of simulations in which the expansion
for the self-gravity of the halo particles is carried to increasing
azimuthal order.  As in the other figures, the dashed line shows the
mass profile from the function eq.~(\ref{Hernquist}). The final mass
profiles for $l_{\rm max} = 2$, 4, 8, 12 \& 16 are barely
distinguishable. All these experiments are for the case of a short bar
with $a=0.2r_s$.}
\label{lmax}
\end{figure}

\begin{figure}[t]
\includegraphics[width=\hsize]{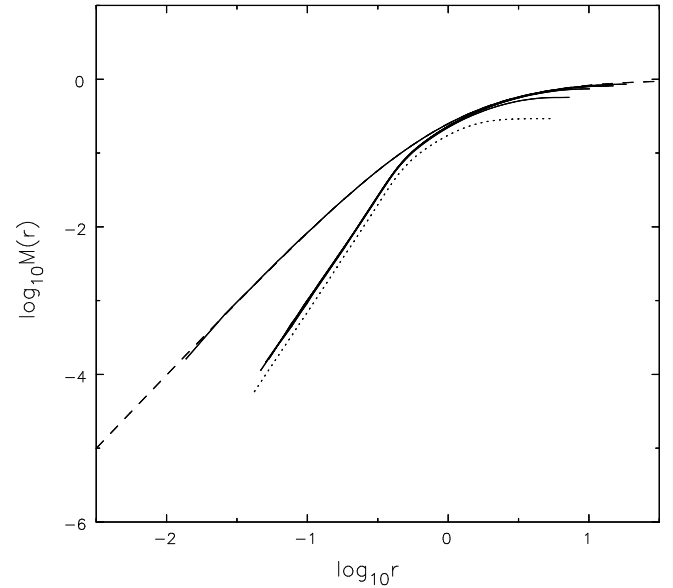}
\caption{\footnotesize The initial and final mass profiles in a series
of simulations of the fiducial run, but in which the truncation radius
of the halo, $r_{\rm cut}$, is varied.  The dashed line shows the mass
profile of the theoretical Hernquist halo and the dotted curve shows
the final profile only in the extreme case of $r_{\rm cut}=2$.}
\label{cutoff}
\end{figure}

\subsection{Other checks}
The code I have used tabulates coefficients of the surface harmonic
expansion of the interior and exterior masses on a radial grid for
almost all experiments.  The mass profiles in experiments in which the
number of radial grid points and the rule for their spacing were
varied, yielded results that could hardly be distinguished from those
with the standard values (middle row, Fig.~\ref{converge}).
Furthermore, results from experiments in which the time step was
halved, and the multi-zone time step scheme (Sellwood 1985) was turned
off, overlay those with the standard step and integration scheme
almost perfectly.  As noted above, other tests revealed that the
outcome was insensitive to the growth-time of the bar over a broad
range of values.

These simulations are heavily smoothed, in the sense that only
low-order multipoles ($l\leq4$, $l\neq1$) contribute to the
self-gravity of the particles.  I have therefore tried increasing
$l_{\rm max}$ to 8, 12 \& 16, with no noticeable effect, even for a
short bar, as shown in Figure~\ref{lmax}.  The same plot includes a
curve with $l_{\rm max} =2$, which is barely distinguishable from the
others.  These experiments include both even- and odd-$l$ terms,
except $l=1$ is always turned off.

Figure~\ref{cutoff} shows that the Hernquist halo can be truncated for
any $r_{\rm cut} \ge 5r_s$ with only a slight effect on the change to
the inner mass profile.  Setting $r_{\rm cut} = 2r_s$ (dotted curve)
significantly decreases the unperturbed density everywhere, including
in the cusp, although the density change is not very different.
However, the benefit of severe truncation, in terms of putting more
particles in the dynamically important region, is modest; merely $\sim
43\%$ of the full Hernquist halo is discarded with the severe
truncation of $r_{\rm cut} = 5r_s$.  Truncating the more extended NFW
mass profile is more beneficial in this regard, however.

These tests have shown that results from these experiments with rigid
bars are insensitive to all numerical parameters, and do not change
when a field method is substituted for the grid to determine the
gravitational forces from the particles.  While the behavior of
simulations using other $N$-body methods has not been tested here,
results from the different test of several methods presented by
Hernquist \& Barnes (1990) suggest that the performance of other
methods may not be radically different.

\section{Behavior at resonances}
\label{resonances}
The stark contrast between the predictions of WK07a and the robust
behavior of my simulations requires explanation.  Since their analysis
focuses on resonances, I here examine the resonant interactions in my
simulations.

\begin{figure}[t]
\includegraphics[width=\hsize]{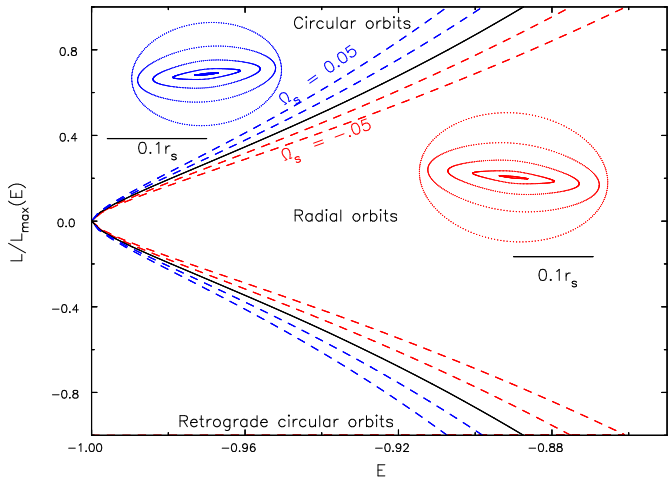}
\caption{\footnotesize The solid curve shows the locus of the ILR in
the space of energy and fraction of the maximum angular momentum for
$\Omega_b = 0.5$ near the center of a Hernquist halo.  The resonance
extends to retrograde orbits in which the signs of $\Omega_\phi$ and
$l$ are reversed.  The dashed curves show the loci of orbits that are
not precisely resonant, and precess at the rates $\Omega_s = \pm
0.025$ and $\Omega_s = \pm 0.05$ relative to the pattern.  The closed
orbits shown are representative of those that precess at $\Omega_s =
\pm 0.05$ relative to the disturbance; they have a wide range of
sizes, with the more eccentric orbits being smaller.  The horizontal
lines, which have a length of $0.1r_s$, show the linear scale for the
orbits.}
\label{locus}
\end{figure}

\subsection{Inner Lindblad Resonance}
As Weinberg and his collaborators have reported, I find that the inner
Lindblad resonance (ILR) is the most important in the early stages of
these particular experiments with massive, skinny bars.  In Paper I, I
found that corotation and the direct radial resonance were the two
most important resonances when using more realistic bars in
simulations that evolved on a much longer timescale and produced
little density change.\footnote{The direct radial resonance arises
when the period of radial motion of a particle is equal to the bar
rotation period; interactions at this resonance can be strong only for
particles on near polar orbits.}  The relative importance of the
different resonances in individual cases depends on the radial
variation of the quadrupole field strength and the density of
particles as functions of the actions, as described in WK07a.

The solid curve in Figure~\ref{locus} shows the locus of the ILR in
the space of energy and fraction of the maximum angular momentum
$L_{\rm max}$ for a quadrupole perturbation with $\Omega_b = 0.5$ in
the Hernquist halo.  The range of $E$ shown is strongly restricted to
the part deep in the center of the potential.  The condition for the
resonance is $\Omega_b = \Omega_\phi - \Omega_r/2$, where $\Omega_r$
and $\Omega_\phi$ are respectively the uniform angular frequencies of
the radial and azimuthal motion of the particles (Binney \& Tremaine
1987, hereafter BT87, ch2).  The solid curve in the figure shows that
more eccentric resonant orbits are more tightly bound (have lower $E$)
than more nearly circular orbits.  The lower half of this Figure shows
the similar resonance for retrograde orbits for which $\Omega_b =
\Omega_\phi + \Omega_r/2$, with $\Omega_\phi$ negative.

As described in BT87 (ch 6) orbits at the ILR drawn in a frame that
rotates with the perturbation are stationary ellipses.  Lynden-Bell
(1979) pointed out that one can regard nearly resonant orbits as
pursuing ellipses that precess relative to the pattern at the slow
angular rate $\Omega_s = \Omega_b - (\Omega_\phi - \Omega_r/2)$.  The
dashed curves in Fig.~\ref{locus} show the loci of lines of constant
$\Omega_s$ along which all orbits precess at the same slow rate
relative to the pattern.

The sign of the average angular momentum exchange between nearly
resonant orbits and the perturbation is determined by their relative
precession rate.  Orbits with small positive $\Omega_s$ gain $L$ on
average, while those with negative $\Omega_s$ lose on average; the net
effect at the resonance depends on the relative numbers of gainers and
losers, which depends on the gradient of the particle density in
frequency across that resonance.\footnote{The evolution of the pattern
speed in these models is rapid, in contrast to the slow trapping of
orbits discussed by Lynden-Bell (1979).}

\subsection{Coverage}
In order to show that the simulations are capturing the resonant
behavior properly, Holley-Bockelmann \etal\ (2005) and WK07b determine
the difference between the density of particles at two different times
in the space of the two integrals $E$ and $L$ (more precisely
$L/L_{\rm max}(E)$).  They evaluate the density in this space from the
particles in the simulation using a smoothing kernel, and color code
regions by the change in density between the two times.  They also
draw the loci of several resonances and call attention to the changes
associated with resonances.

Their diagnostic therefore requires phase space to be so densely
populated that the appropriate change in density occurs at every point
in the 2-D space of these integrals, which requires many particles at
each point and a very large number in total.  However, the resonance
extends over a long path through this space, and it is unnecessarily
stringent to insist that the correct balance between gainers and
losers be fulfilled separately at each point.  Instead, the balance
need be realized for all resonant particles, which requires many times
fewer particles.

\begin{figure*}[t]
\includegraphics[width=.5\hsize]{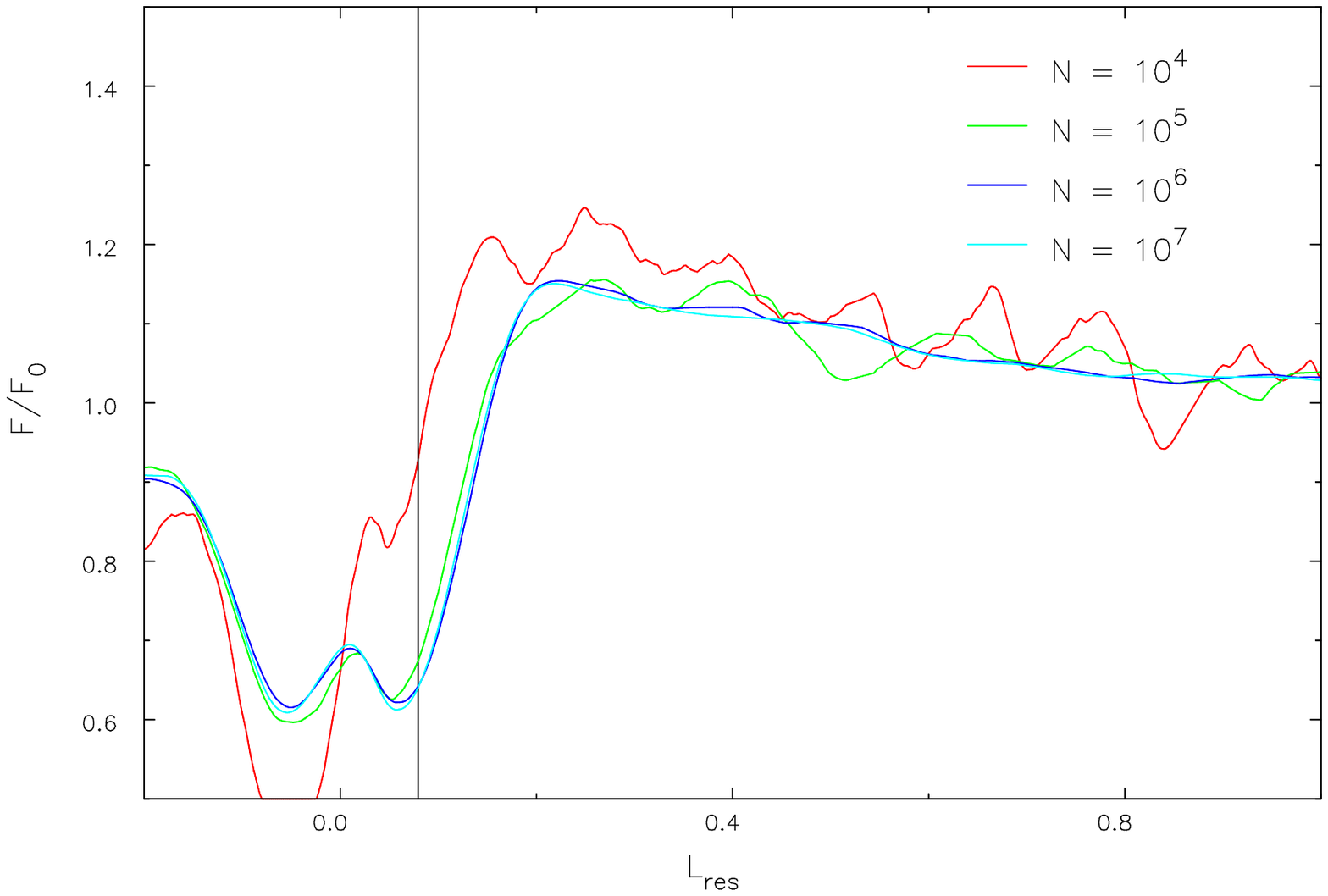}
\includegraphics[width=.5\hsize]{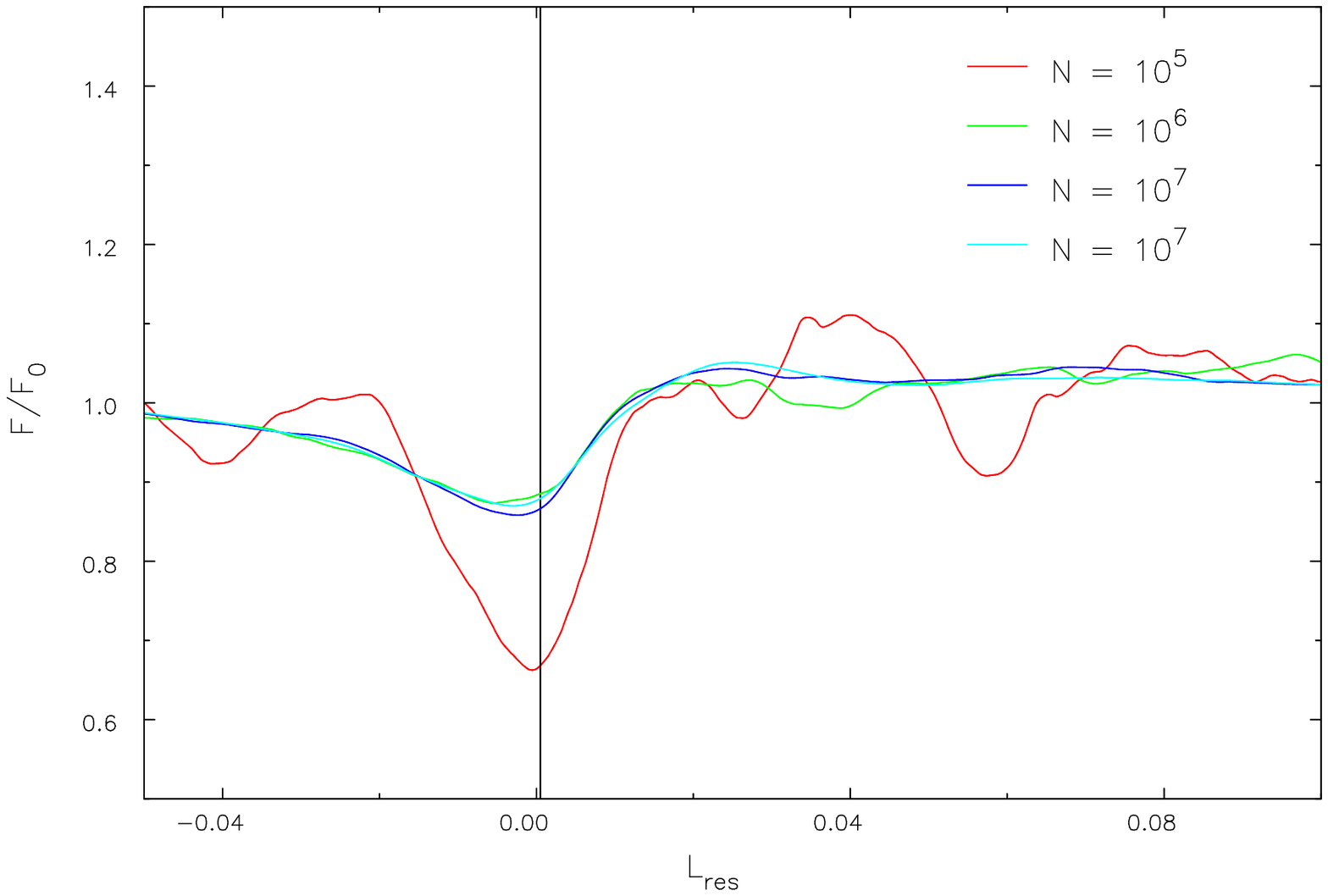} \\
\includegraphics[width=.5\hsize]{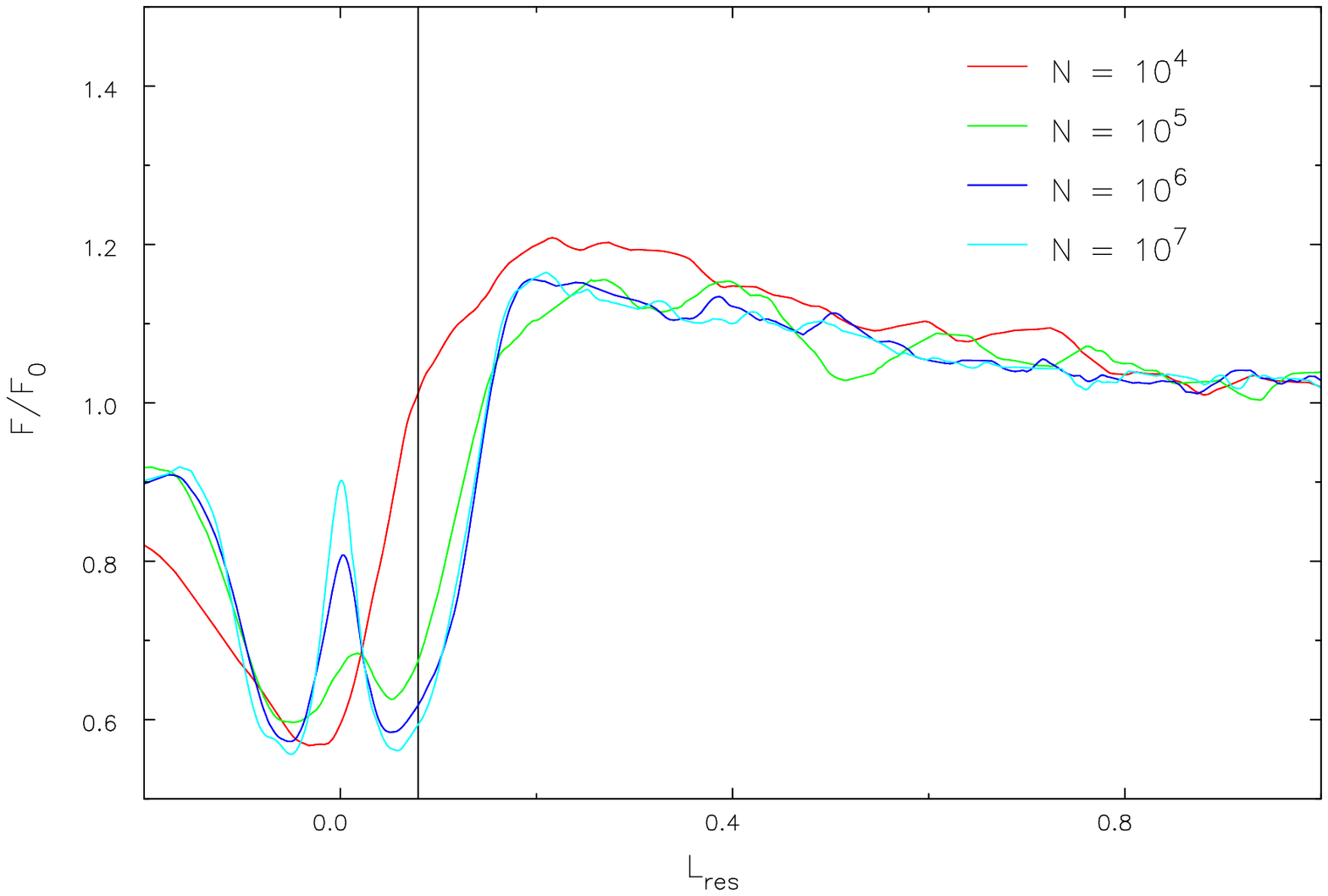}
\includegraphics[width=.5\hsize]{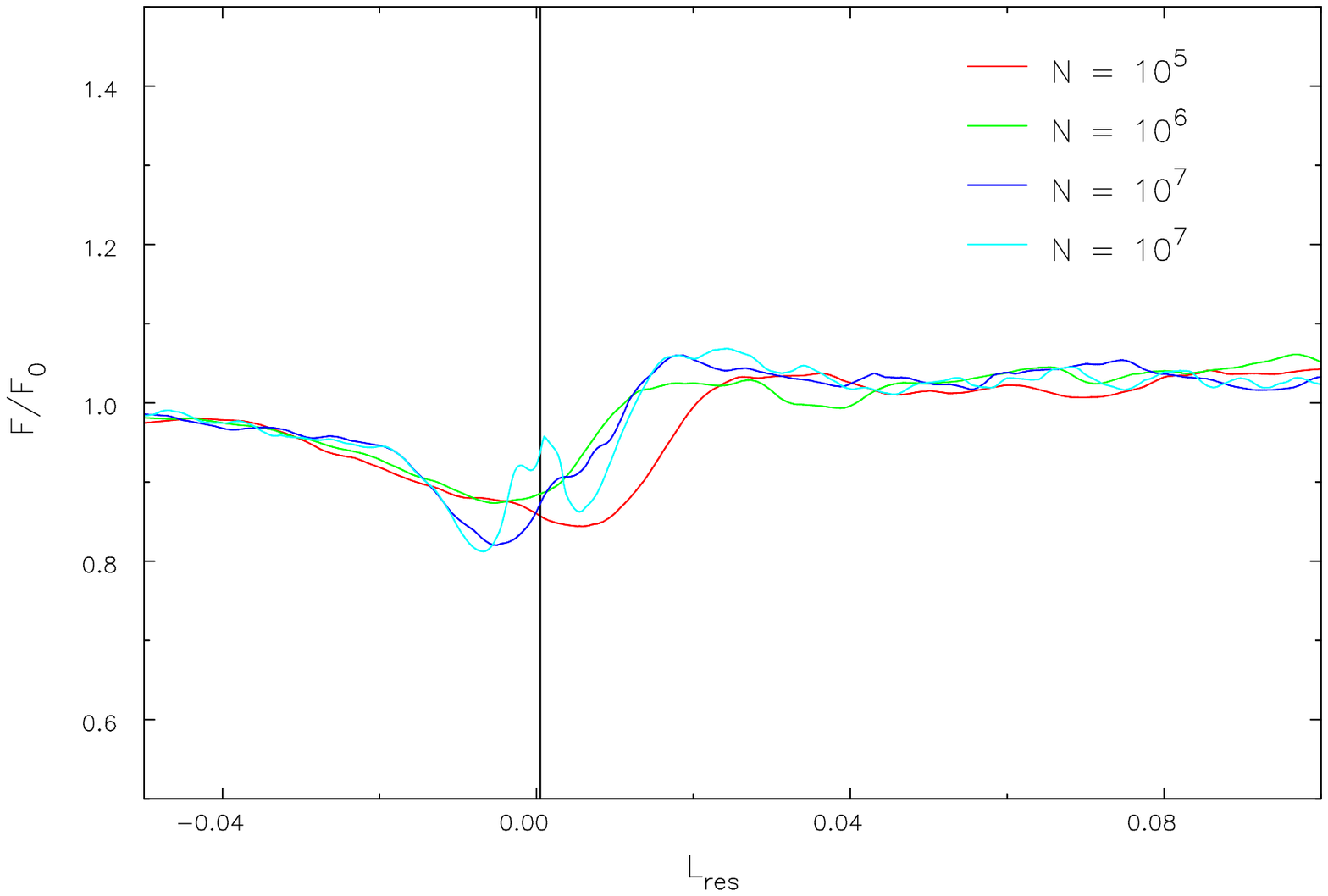}
\caption{\footnotesize The ratio $F(L_{\rm res})(t)/F(L_{\rm res})(0)$
for the ILR for two convergence sequences shown in
Fig.~\ref{converge}.  The left panels are for $t=8$ for the long bar
($a=r_s$) while the right panels are for $t=4$ in the cases with
$a=r_s/5$.  The top panels employ a fixed kernel width for all cases,
while the kernel width is reduced as $N$ rises in the bottom panels.
The vertical lines mark the instantaneous position of the ILR in each
case, which is $L_{\rm res} \sim 5 \times 10^{-4}$ for the rapidly
rotating short bar.  The cyan line for the short bar case is for
unequal mass particles, particles have equal mass in all other cases.}
\label{Lres}
\end{figure*}

\subsection{A Superior Diagnostic}
To demonstrate that the appropriate resonant exchanges are occurring
at much lower particle numbers than WK07a suggest are needed, I
compute the average density change along lines of constant frequency
difference $\Omega_s$, such as the dashed lines in Fig.~\ref{locus}.
The average so defined is a function of the single variable
$\Omega_s$, but since this is not an intuitive quantity, I map
$\Omega_s$ to the quantity $L_{\rm res}$, the angular momentum of the
circular orbit that precesses at the rate $\Omega_s$ relative to the
perturbation.

In practice, I compute the frequencies $\Omega_r$ and $\Omega_\phi$
for every particle in the simulation in the spherically averaged
potential at some moment during the evolution.  I compute the
frequency difference $\Omega_s$ for a selected resonance and evaluate
the density of particles at each $\Omega_s$ using a 1-D kernel
estimate.  Then the relation between $\Omega_s$ and $L_{\rm res}$
yields the 1-D function $F(L_{\rm res})$ at the selected time (Paper
I).  This diagnostic is therefore both easier to show and less
affected by shot noise than is the density of particles as a function
of the two classical integrals $E$ and $L$.

Once the halo density profile starts to change in these experiments,
the spherically averaged gravitational potential and the resonant
locus also change.  I therefore focus here on the early stages before
this complication becomes important, although $F(L_{\rm res})$ can be
computed with a little more effort for any arbitrary potential, as
shown in Sellwood \& Debattista (2006).

Figure~\ref{Lres} shows the ratio of $F(L_{\rm res})$ to its initial
value for the ILR in the convergence tests shown in the top and middle
rows of Fig.~\ref{converge}.  The quantity shown is the ratio of
$F(L_{\rm res})$ to its undisturbed value for different values of $N$.
The left panels show results at $t=8$ for the long bar ($a=r_s$) and
the right panels at $t=4$ for the short bar ($a=r_s/5$).  The upper
panels show the results with a fixed kernel width, while the width of
the smoothing kernel is halved for every factor 10 increase in $N$ in
the lower panels.  The cyan curve in the right panels is for unequal
mass particles with a further reduction of the smoothing kernel width
in the lower panel.

As $N$ is increased by three orders of magnitude in the large bar case
(left panels), the results quickly converge when a fixed kernel is
employed (upper).  Reducing the kernel width as $N$ rises (lower)
reveals more detail of the function shape.  Even for the smallest
particle number ($N=10^4$), the function shows a substantial change
associated with the resonance, but lacks the central spike at $L_{\rm
res}=0$ visible in the other cases.  The local maximum at $L_{\rm res}
= 0$ arises because particles of very low angular momentum have orbits
that precess at such a high rate they are well inside the ILR and
their angular momenta are little affected by the perturbation.  The
kernel width is too large to reveal this feature in the $N=10^4$
case.

Results for the short bar are shown on the right, which again quickly
converge with a fixed kernel width (upper).  When the kernel width is
decreased (lower), a central spike appears only in the case of unequal
mass particles (cyan line), for which I also refined the radial grid
to place more shells in the inner parts.  Note that the resonance is
still well-populated in the other three experiments, since $F(L_{\rm
res})$ is strongly affected in the appropriate sense, and the time
evolution of the pattern speed and density profile, shown by the
dotted curves in the middle row of Fig.~\ref{converge}, are no
different from those in the coarser experiments.  The number of equal
mass particles above the resonance in these cases is too small to
reveal the spike, whereas the unequal mass case packs in many high
frequency particles; clearly, adding particles that are adiabatically
invariant to the perturbation can have no effect on the outcome.

\section{Discussion}
\label{reasons}
\subsection{Frequency broadening}
The range of $L_{\rm res}$ over which the ratio departs from unity in
Fig.~\ref{Lres} indicates the extent of the resonance during this
short time interval, and one can count the numbers of particles within
this range.  For the large bar, in the left panels of Fig.~\ref{Lres},
I find fully 10\% of equal-mass particles have $\Omega_s$ within the
range affected by the resonance, but this factor drops to $\ga 0.7\%$
for the shorter bar ($a = r_s/5$).  While this smaller fraction
clearly implies that a larger number of particles is needed in this
more delicate case, as already found empirically in
Fig.~\ref{converge}, the $\sim 7,000$ resonant particles in a
simulation with $N=10^6$ are enough to capture the appropriate
response.  The resonant fraction with unequal particle masses rises to
20\%, even in this short bar case, but the evolution is no different.

The fraction of particles that participate in the ILR is far higher
than expected in the calculations by WK07a.  This is because their
estimate of the resonance width neglects frequency broadening due to
the time evolution of the perturbation.  Resonances are broadened both
by the time evolution of the amplitude, which rises smoothly from zero
to its full value in 10 time units, and also because the bar slows
over an even shorter time interval (see Fig.~\ref{converge}).  The
initial bar period of the large bar is $\sim 12.5$ time units, which
is longer than both the turn-on time and the slow-down time.  The bar
period for the shorter bars is $\sim 3$ time units initially, and
therefore frequency broadening of resonances is slightly less than for
the large bars, but is still highly significant.

Thus the estimates from WK07a of the particle numbers required to
``cover'' the resonance are for very slowly evolving perturbations,
and not for realistic experiments that might produce a large density
change.  It is perplexing that these authors explicitly discount
frequency broadening, since Weinberg (2004) has already shown that the
pattern speed evolution depends upon the time history of the
perturbation -- a clear indication that the resonant interactions
responsible for friction do depend upon the broadening of the
resonance by the time evolution of the perturbation.

\subsection{Particle noise}
\label{noise}
As for the coverage issue, the timescale for bar pattern speed
evolution is so rapid in cases in which the halo density is changed
that the time-scale for interactions with the bar is very short.
Questions of orbit quality in a noisy potential seem of marginal
relevance when the location of the resonance moves faster than any
reasonable orbit diffusion rate.

From a crude analysis, BT87 find the relaxation time of a collection
of $N$ point masses is
\begin{equation}
\tau_{\rm relax} \simeq {0.1N \over \ln N} \tau_{\rm orb},
\end{equation}
where $\tau_{\rm orb}$ is a typical orbit period.  This is an
underestimate of the relaxation time for most collisionless $N$-body
methods, which smooth the gravitational field through particle
softening, limited mesh resolution, or some form of filtering of the
high-spatial frequencies of the potential.  To be conservative, I
ignore this favorable caveat for now, and continue the argument with
the above crude estimate of the relaxation rate.

Since the relaxation time is approximately the time to cause an order
unity change to the initial energy of a typical particle, the
fractional energy change per orbit $\Delta E/E \sim 10 \ln N/N$.  If
fractional changes in the important frequencies of an orbit scale as
the fractional change in energy of the orbit (this appears to be
approximately true in many potentials), orbit scattering will be too
slow to affect resonant interactions as long as the fractional change
in the bar frequency in one orbit
\begin{equation}
\left| {\Delta \Omega_b \over \Omega_b} \right| \gg {10 \ln N \over N}.
\end{equation}
This very crude estimate suggests that relaxation is utterly
irrelevant when $|\Delta \Omega_b / \Omega_b| \sim 1$ in one orbit
(\eg, Fig.~\ref{converge}), and will not become an issue except for
very mildly braked bars simulated with small numbers of particles.
Again WK07a conclude that much larger numbers of particles are needed
because their analysis fails to take the changing pattern speed of the
perturbation into account.

\subsection{Self-gravity methods}
\label{discmeth}
Relaxation is conventionally thought of as the cumulative effect of
pair-wise encounters between particles, as above, but it can also be
regarded as the effect of square root of $N$ type excitations of a
number of neutral modes of the equilibrium system, as remarked by
Sellwood (1987) and calculated by Weinberg (2001).  WK07a attempt to
separate $N$-body fluctuations into small- and large-scale noise and
appear to associate simple 2-body scattering with small-scale noise
and large-scale noise with that from the neutral modes excited by shot
noise in the particle distribution (Weinberg 2001).  Such a
distinction is artificial, since both approaches describe the same
physical process.

Hernquist \& Barnes (1990) and Hernquist \& Ostriker (1992) measured
very similar relaxation rates in spherical models simulated by various
$N$-body methods.  Their important finding can be understood from
either approach.  First, the Coulomb logarithm appears in the
expression for the relaxation rate because every decade of impact
parameters contributes equally (BT87).  As collisionless $N$-body
methods have a limited range of spatial resolution, the number of
decades over which scattering must be integrated is strictly limited,
and not very different from method to method.  Second, only a limited
number of neutral modes affect the behavior of an $N$-body simulation,
because softening, grid resolution, or truncation of the field
expansion quickly cuts off the dynamical influence of the higher modes
that have shorter wavelength.  Thus either conceptual approach to the
influence of noise leads to the same conclusion that no valid $N$-body
method is dramatically less collisional than any other (Hernquist \&
Barnes 1990).  (Methods that do not employ many particles per
effective softening volume should manifest higher relaxation rates.)

WK07a argue correctly that a well-chosen basis allows forces from
unwanted fluctuations on small spatial scales to be filtered out.
Despite this, Hernquist \& Ostriker (1992) found little improvement in
the relaxation rate from this method over others.  Thus we conclude
that all methods filter out all but the longest range encounters,
unless resolution is taken to extreme, leading to at most marginal
differences of quality between different methods.

\subsection{Previous work}
The ability of $N$-body simulations to capture resonant exchanges with
a perturbation has previously been demonstrated in the case of disk
instabilities.  Global modes that lead to bars rely on the emission of
angular momentum at the ILR and its absorption at other resonances
farther out in the disk (\eg, Kalnajs 1977).  Since the mode is driven
by 2nd-order coupling between the particles and the wave at
resonances, the dynamics resembles that of bar-halo coupling in 3-D.
In particular, the third action for each orbit is zero for precisely
spherical potentials, making the unperturbed motion of each halo
particle no more complicated than in 2-D.  It is worth noting that
Rybicki (1972) pointed out that 2-D disks are essentially more
collisional than 3-D systems, which would argue that if relaxation
were an important problem, it ought to be {\it harder\/} to get disks
right.

Sellwood (1983), Sellwood \& Athanassoula (1986) and Earn \& Sellwood
(1995) report they are able to reproduce the global bar modes of some
disks in simulations with comparatively modest numbers of particles.
Tests of a disk without velocity dispersion, employing a large
softening length to inhibit local instabilities, may not be a fair
comparison with 3-D systems.  However, Earn \& Sellwood (1995) present
results for disks with velocity dispersion using both a field method
and a 2-D polar grid.  The predicted eigenfrequency was reproduced to
within 5\% percent using a field method with as few as 15K particles,
and agreement with theory improved for moderately larger $N$.  Results
with the polar grid were discrepant because gravity softening was
required, but Fig.~4 of that paper shows that the trend with softening
length could plausibly extrapolate back to the predicted frequency at
zero softening.

These reassuring results indicate that simulations do indeed capture
the appropriate collective response at resonances, without requiring
vast numbers of particles.  Again the dynamical response of the
collection of particles extends over the entire resonance, broadened
by the growth-rate of the bar, and does not need to reproduce the
detailed balance of gainers and losers at every point in integral
space.

\subsection{Numerical convergence}
WK07a argue that numerical convergence alone is not a guarantee that
the result is correct.  They suggest that low-$N$ experiments could
converge to the wrong result, where friction is determined by one-time
encounters between the particles and the bar, while the proper
resonant behavior would not be revealed until some much larger
particle number is reached.

This argument is unconvincing for several reasons.  First, if coverage
were inadequate, as WK07a note, the exchange of angular momentum with
the bar would depend on just the few particles that happened to be in
resonance, resulting in significant stochasticity in the evolution.
The pattern speed and density changes would depend on the random seed,
which I do not observe, and the curves in Fig.~\ref{converge} could
not overlay so perfectly.  Second, as shown in Fig.~\ref{Lres}, I have
been able to detect the influence of resonances over a wide range of
$N$.  Third, WK07b estimate that $N \ga 10^8$ equal mass particles
should be sufficient for a strong bar with semi-major axis equal to
the profile break radius $r_s$.  I have presented a result with $N =
1.6 \times 10^8$ that behaves no differently from experiments with
much lower $N$.  This sequence of experiments therefore demonstrates
that nothing different occurs when their criteria are met.

WK07b present a result for a strong bar with length equal to $r_s/6$
in which the evolution differs when $N$ is increased from $10^6$ to $5
\times 10^6$.  I have been unable to reproduce a change in behavior at
any $N$ in tests with similar, though not identical, bars; their bar
had an axis ratio of 5:1, which I have also used, but since it is
possible the quadrupole field of their bar is weaker than given by
eq.~(\ref{quadrup}), I chose to present a 4:1 bar in the bottom row of
Fig.~\ref{converge} (as described in \S4.1).  It is unclear why WK07b
find a different result with different $N$, but my failure to observe
differences of this kind in a similar regime suggests that the
difference they report must be due to factors other than they suggest.

\section{Halo density reduction}
\label{physics}
\subsection{Cusp flattening}
\begin{figure}[t]
\plotone{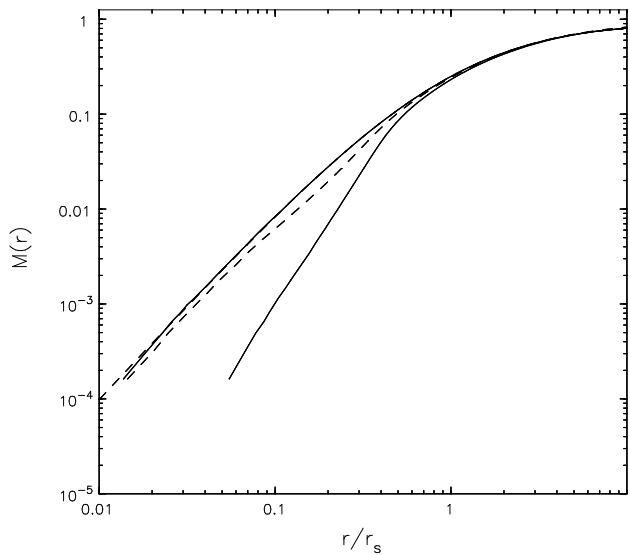}
\caption{\footnotesize The changes to the mass profile for the
fiducial bar with the monopole term active (solid curves) and in an
identical case when the $l=0$ term of the halo mass distribution is
held fixed (dashed).  These experiments employed $10^7$ unequal mass
particles.}
\label{rigidh}
\end{figure}

Figure~\ref{rigidh} compares the mass evolution in the fiducial run,
for unequal mass particles, with that when the monopole term of the
halo mass distribution is held fixed.  It should be noted that these
two runs differ only in the monopole terms, the gravitational field
from the $2 \leq l \leq 4$ density response of the particles is
included in both cases.  It is clear that including the change in the
potential that arises from the change to the radial mass profile is
crucial for creating a large density reduction, as previously found
for driven bars (Sellwood 2003).

Thus flattening of the cusp is a collective effect that is suppressed
when the self-consistent potential changes are eliminated. Once the
collective change is initiated, the different radial mass profile
allows somewhat more angular momentum to be accepted by the resonant
particles; the torque in the self-consistent case is some 20\% larger
at its peak, near $t=8$, than when the central attraction is held
fixed.  This is physically reasonable, since adjustments to the
central attraction of the mass distribution will further broaden the
resonances.  Note that the self-consistent density change could not be
predicted from simple perturbation theory, since the global potential
in which the particles move undergoes substantial evolution on an
orbital time-scale during the cusp-flattening stage (see
Fig.~\ref{mrplot}).

WK07b report much smaller density reductions than I find with similar
strong, skinny bars.  It is likely that the collective effect I find
to be responsible for large density reductions is inhibited by the
rigid mass component they include.  To test this hypothesis, I have
tried similar experiments that include the rigid bar monopole term and
find that density reduction is almost entirely suppressed.

\begin{figure}[t]
\includegraphics[width=\hsize]{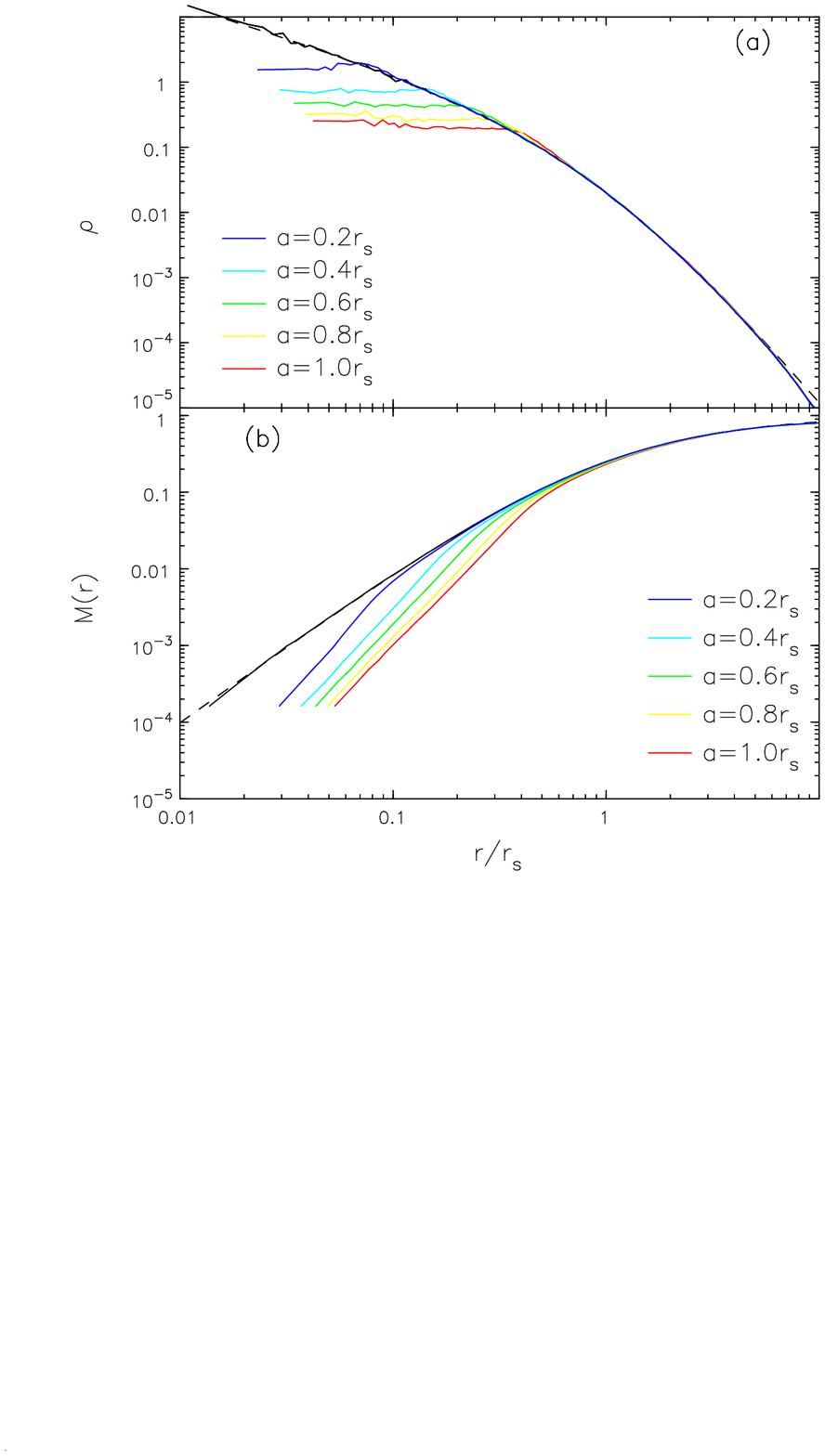}
\caption{\footnotesize Results from five different experiments with
different bar lengths.  (a) The dashed line shows the initial profile
given by eq.~\ref{Hernquist}.  The solid lines show estimates from the
particles of the initial (black) and final (color) density profiles
from a series of runs with different bar semi-major axes, $a$.  (b)
The same results plotted as halo mass enclosed as a function of
radius.}
\label{barlength}
\end{figure}

\subsection{Variation of bar properties}
Here I report the results of changing the physical parameters of the
bar perturbation: its length, mass, and axis ratio.

Figure~\ref{barlength} shows the final density profiles from a series
of five separate simulations using bars of different lengths.  The
lengths span the range $0.2 \leq a \leq 1$, in equal steps of $\Delta
a = 0.2$, while the nominal bar axis ratios are kept at $a/b = 5$ and
$a/c = 10$.  As above, the bar mass is half the enclosed halo mass
at $r=a$ and the initial pattern speed places corotation at the bar
end.  In all experiments shown in Fig.~\ref{barlength}, the final halo
density is flattened inside $r \simeq 0.3a$, while remaining
essentially unchanged at larger radii.

\begin{figure}[t]
\includegraphics[width=\hsize]{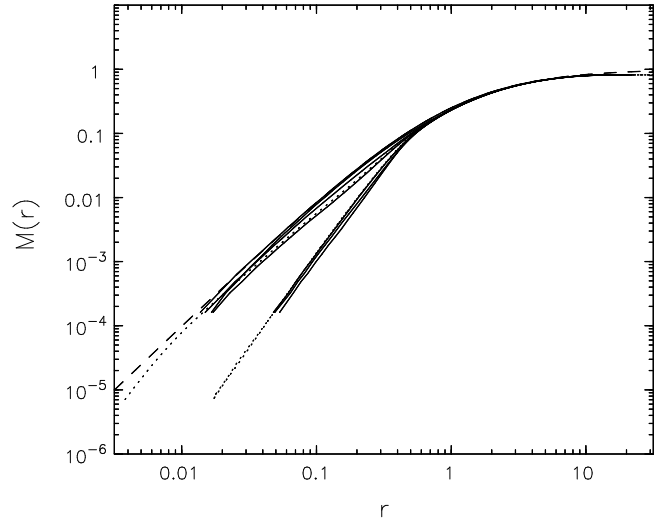}
\caption{\footnotesize Solid lines show the initial and final mass
profiles from a series of runs in which the bar axis ratio $b/a$ was
varied.  The pronounced gap in the final mass profiles is bracketed by
two runs in which $b/a = 0.32$ and $b/a = 0.30$; the dotted lines show
reruns of these models with more individual-mass particles.  The
dashed line shows the Hernquist profile.}
\label{axrat}
\end{figure}

\begin{figure}[t]
\includegraphics[width=\hsize]{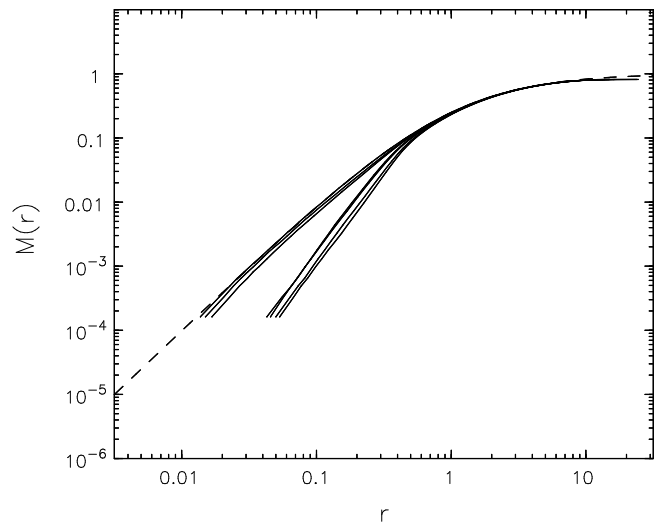}
\caption{\footnotesize As for Fig.~\ref{axrat}, but from a series of
runs in which the bar mass was varied.  The sharp transition occurs
between $0.0625 \leq M_b \leq 0.07$.}
\label{bmass}
\end{figure}

\begin{figure}[t]
\plotone{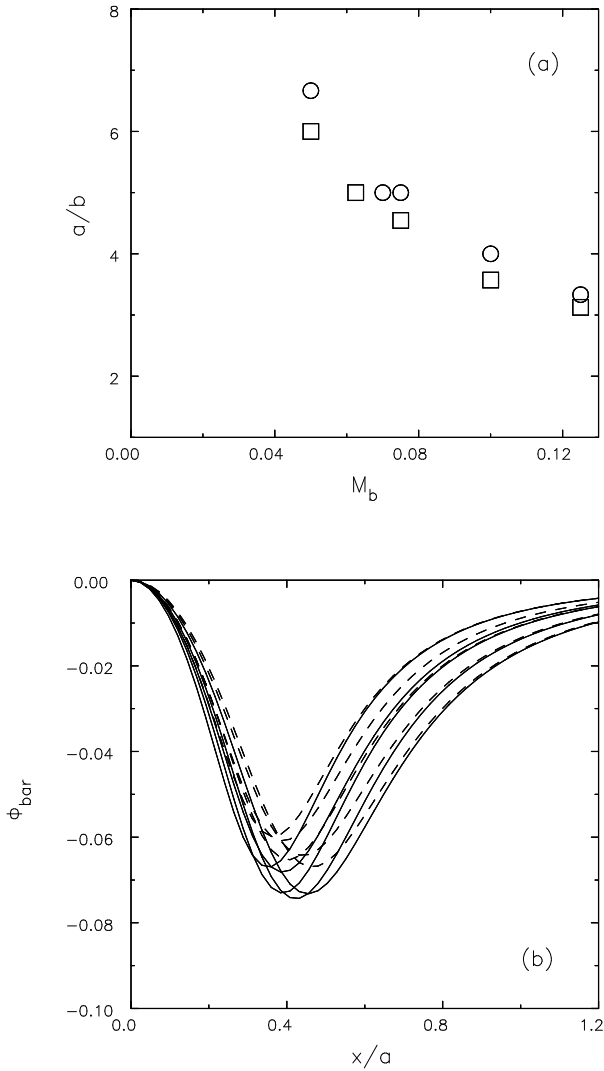}
\caption{\footnotesize Results near the sharp transition.  (a) Circles
indicate the cusp flattened for the bar mass and axis ratio, while
squares indicate it did not.  The bar semi-major axis is held constant
at $a=r_s$ in all cases.  (b) The quadrupole potential along the bar
major axis given by eq.~(\ref{quadrup}) for the bars in the
simulations shown in (a).  Bar potentials that caused the cusp to
flatten are drawn with solid lines, those that did not are dashed.}
\label{qsharp}
\end{figure}

Figure~\ref{axrat} shows the effect of changing the bar axis ratio
$b/a$.  The nominal bar axis ratios in the models shown range from
$a/b = 5$ to $a/b=2$; in all cases, $a=r_s$, $M_b = 0.125$, and
$\Omega_b = 0.5$ initially.  The more elliptical bars produce large
density changes, whereas rounder bars have little effect.  A {\it
sharp transition\/} is evident in these results between $0.30 < b/a <
0.32$.  Tests with the SCF method (Hernquist \& Ostriker 1992) also
reveal the sharp transition at the same bar axis ratio.

A similar effect is seen in Figure~\ref{bmass}, in which $0.050 \leq
M_b \leq 0.125$, \ie\ the bar mass ranges from 20\% to 50\% of the
enclosed halo mass, while the bar axis ratio is held fixed at
$b/a=0.2$.  The sharp transition occurs between $0.0625 \leq M_b \leq
0.07$.

\subsection{Sharp transition}
\label{bimodal}
The bimodal nature of the density change shown in Figs.~\ref{axrat} \&
\ref{bmass} appears to be real.  The models evolve more slowly as
friction is weakened by reducing the bar quadrupole field, either by
making the bar rounder or by reducing its mass, but the halo density
change undergoes an abrupt transition as the parameter is varied
smoothly.

\begin{figure}[t]
\plottwo{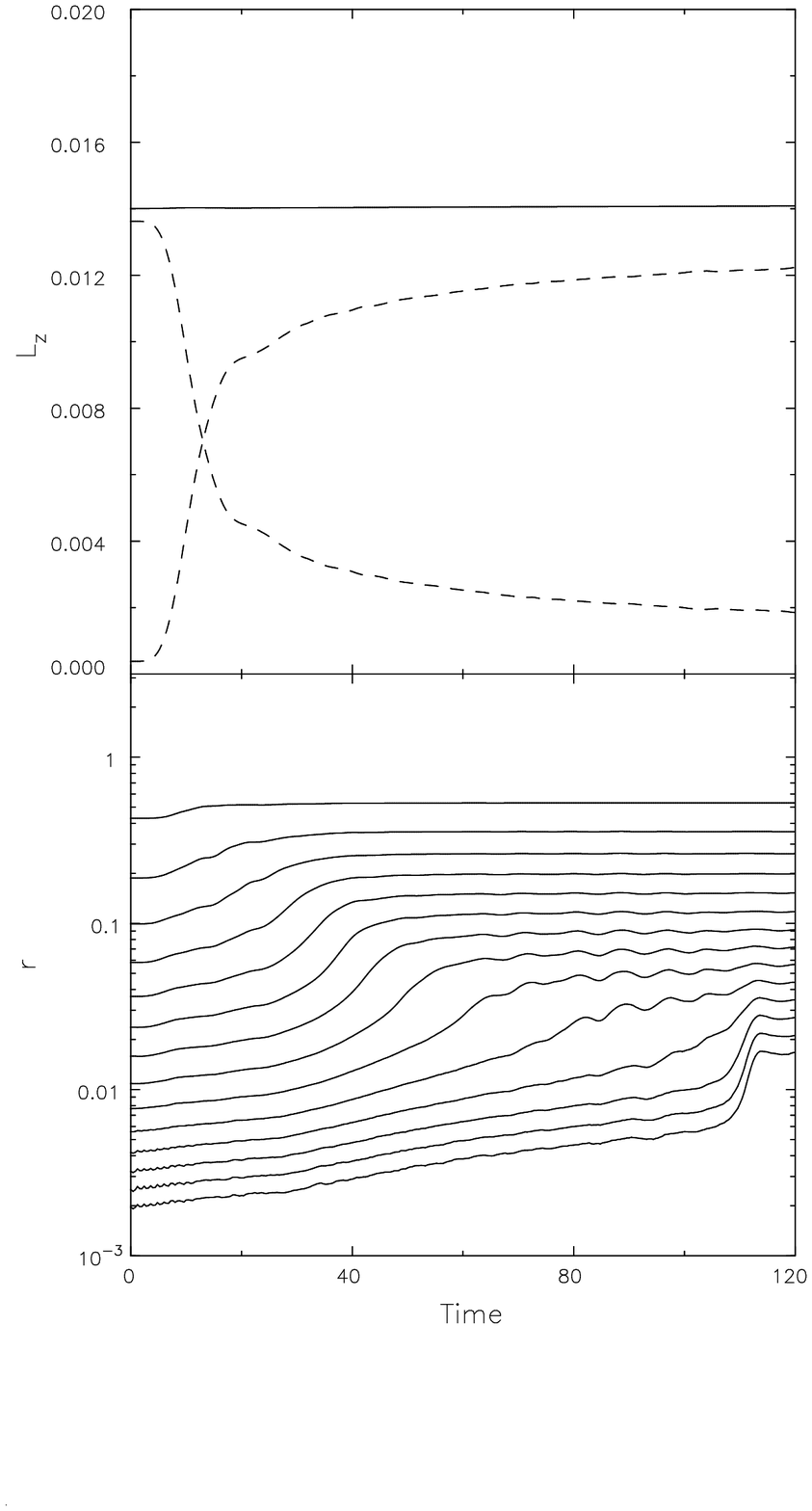}{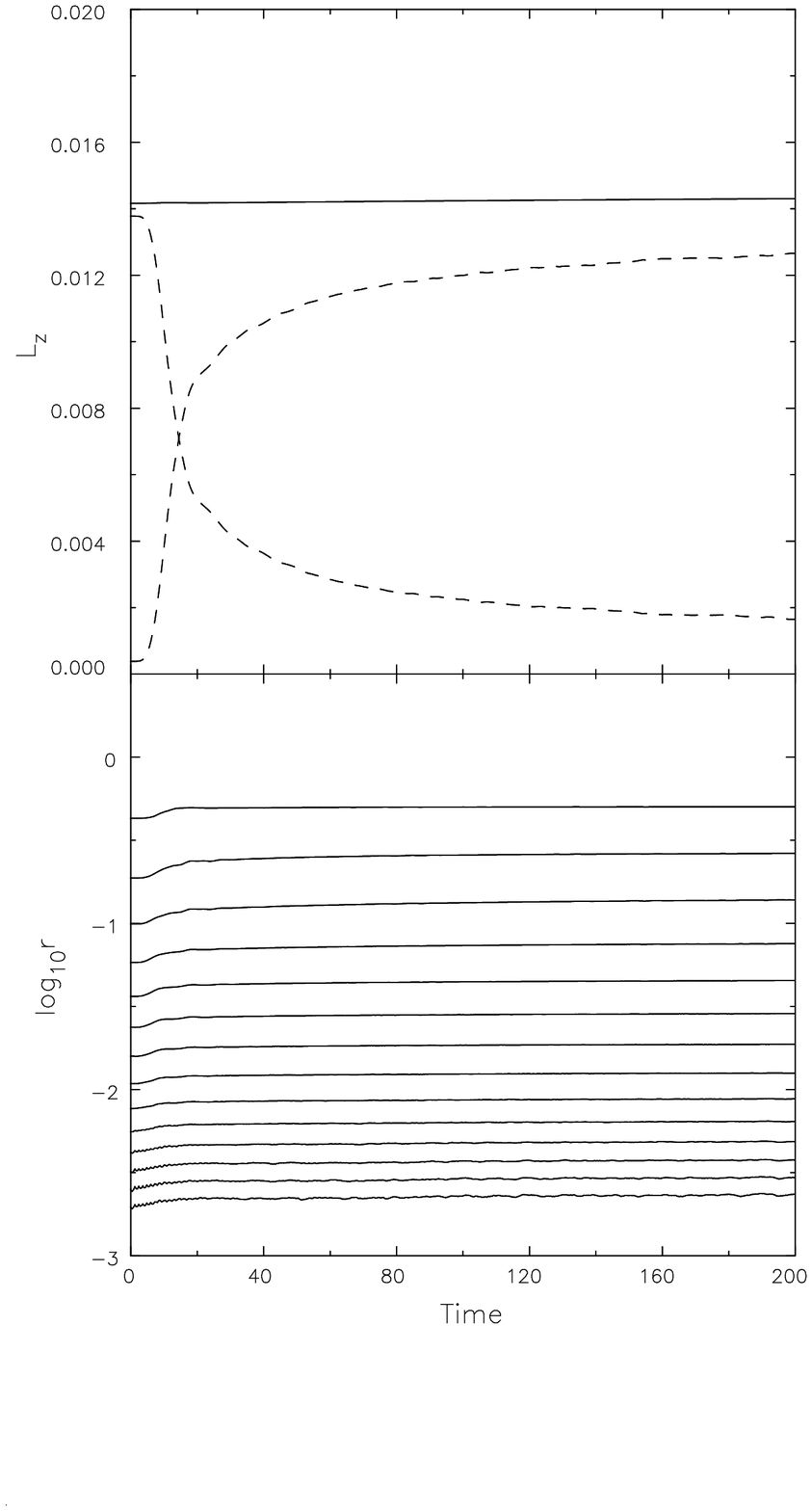}
\caption{\footnotesize Above: The angular momenta of the bar
(decreasing dashed) and of the halo (increasing dashed) and total
(solid) in two simulations straddling the boundary between cusp
flattening and more gradual density change.  That on the left had $b/a
=0.30$ while that shown on the right had $b/a=0.32$.  Below: The
radii containing different fractions of the total number of particles
in the two cases.}
\label{boundary}
\end{figure}

Figure~\ref{qsharp}(a) shows results in the space of the two
parameters $a/b$ and $M_b$ from experiments run to map out the
transition boundary, always for the case of the long bar with $a=r_s$.
The quadrupole fields of the bars are shown in Fig.~\ref{qsharp}(b).
This Figure suggests that there is a critical quadrupole field
strength required to cause the cusp to flatten, which may decrease
slightly towards skinnier bars where the quadrupole peaks at smaller
radii.

It is unclear what triggers the collective response.
Figure~\ref{boundary} shows more information from the two cases that
straddle the sharp transition in the density change as the axis ratio
is changed.  The angular momentum absorbed by the halo (upper panels)
differs very little between the two cases, yet the slightly stronger
bar flattened the cusp at late times (after most of the angular
momentum had been lost) while the other did not.  I have checked that
no dramatic density changes occur in the cases with weaker
quadrupoles, no matter for how long the simulations are continued.
Friction tails off at late times in these runs without producing a
large density change.

Notice also the clear time sequence in the density reduction (lower
left panel); the density in the outer part of the cusp is reduced
before that in the inner part.

The mass profile evolution is insensitive to numerical parameters
everywhere except near the transition.  \S\ref{numtests} presented
numerous tests to show that the evolution of these simulations does
not depend on numerical parameters.  Since WK07b argue that more
delicate cases require larger $N$, I simulated the large, massive bar
model with $b/a=0.32$ with $1.6 \times 10^8$ unequal mass particles,
finding a small change to the mass profile that is no different from
that in simulations with lower $N$.

However, the outcome of experiments for the marginal case of the
large, massive bar with axis ratio $b/a = 0.31$ does depend on
numerical parameters.  In some cases the cusp flattens while in others
it does not; the result is never intermediate, however.  Thus these
simulations cannot pin down the parameter values at which the outcome
changes to better than a few percent.

I have searched the experimental results for a property that could be
the cause of the sharp transition.  I examined resonant exchanges in
two simulations that straddle the boundary using the procedure
described in \S\ref{resonances}, finding only very minor differences
in $F(L_{\rm res})$ between the two cases.  The ILR continues to be
the most important resonance, even at late times when the pattern
speed is about 20\% of its initial value; friction is weak and changes
to $F(L_{\rm res})$ are correspondingly small, but still detectable.
Other properties, such as the amplitude of the bisymmetric distortion
in the halo response, all varied smoothly with the strength of the
quadrupole field.

While the trigger for the collective response that brings about the
large density change remains elusive, further investigation seems
warranted only if a similar sharp transition were found in fully
self-consistent models.

\begin{table}[t]
\caption{Summary of simulations plotted in Figure \ref{ABW}}
\label{ABWval}
\smallskip
\begin{tabular}{@{}lcccccr}
$\Mb$  &   $a$ &   $b$ & $\delta\lambda$ & $\mu$ & fltnd & MoI \\
\hline
\noalign{\smallskip}
0.0139 &   0.2 &   0.04 &   0.00006 &   0.59595 & y &   \\
0.0408 &   0.4 &   0.08 &   0.00033 &   0.25639 & y &   \\
0.0703 &   0.6 &   0.12 &   0.00096 &   0.15082 & y &   \\
0.0988 &   0.8 &   0.16 &   0.00184 &   0.09279 & y &   \\
0.125  &   1.0 &   0.20 &   0.00288 &   0.06936 & y &   \\
0.0139 &   0.2 &   0.04 &   0.00024 &   0.44957 & y & 5 \\
0.125  &   1.0 &   0.20 &   0.00288 &   0.06936 & y &   \\
0.1    &   1.0 &   0.20 &   0.00236 &   0.08386 & y &   \\
0.075  &   1.0 &   0.20 &   0.00176 &   0.11818 & y &   \\
0.07   &   1.0 &   0.20 &   0.00176 &   0.12763 & y &   \\
0.0625 &   1.0 &   0.20 &   0.00156 &   0.60344 & n &   \\
0.050  &   1.0 &   0.20 &   0.00113 &   0.84208 & n &   \\
0.050  &   1.0 &   0.20 &   0.00514 &   0.74350 & n & 5 \\
0.125  &   1.0 &   0.20 &   0.01215 &   0.04334 & y & 5 \\
0.0625 &   1.0 &   0.20 &   0.00629 &   0.10646 & y & 5 \\
0.125  &   1.0 &   0.25 &   0.00283 &   0.07789 & y &   \\
0.125  &   1.0 &   0.27 &   0.00298 &   0.08104 & y &   \\
0.125  &   1.0 &   0.29 &   0.00315 &   0.09306 & y &   \\
0.125  &   1.0 &   0.30 &   0.00323 &   0.09352 & y &   \\
0.125  &   1.0 &   0.31 &   0.00312 &   0.34321 & y &   \\
0.125  &   1.0 &   0.33 &   0.00274 &   0.70371 & n &   \\
0.125  &   1.0 &   0.50 &   0.00296 &   0.91227 & n &   \\
0.125  &   1.0 &   0.33 &   0.00982 &   0.04764 & y & 5 \\
0.125  &   1.0 &   0.50 &   0.00874 &   0.89144 & n & 5 \\
0.125  &   1.0 &   0.50 &   0.01378 &   0.83108 & n & 10 \\
\hline
\end{tabular}
\tablecomments{Columns 1 -- 3 summarize the properties of the bar.
Columns 4 \& 5 give the principal results. Column 6 indicates whether
or not the cusp was flattened and column 7 gives the factor by which
the moment of inertia was increased.}
\end{table}

\subsection{More gradual density changes}
The large density changes emphasized so far are confined to the region
well interior to the end of the bar.  They result from a collective
response of the halo particles to the torque from a massive, skinny
bar.  The perturbing potential is not only stronger than that of the
nominal homogeneous ellipsoidal bar (see Appendix), but is also not
easily related to bars in real galaxies that may have somewhat
different quadrupole fields.  However, it seems unlikely that real
bars, which typically have $a/b \la 3$, are strong enough to provoke
such a collective halo response.

The bars that did not produce large density changes are still strong,
both in mass and in axis ratio.  Friction from these bars does lead to
a slight reduction in halo density over a more extended radial range;
tests reported in Paper I and further tests here confirm that these
results are also insensitive to numerical parameters.  It is likely
that the modest mass profile change reported by Debattista \& Sellwood
(2000), and those discernible in Athanassoula's (2003) results are of
this kind.

\begin{figure}[t]
\includegraphics[width=\hsize]{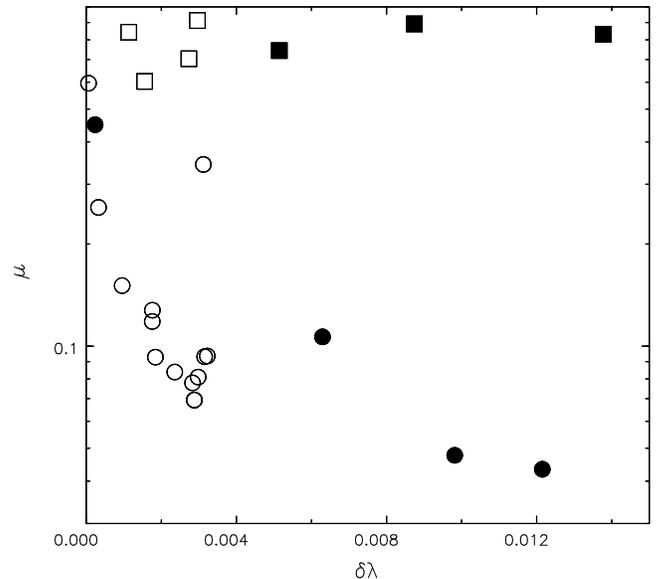}
\caption{\footnotesize Fractional changes, $\mu$, to \Dv2\ in many
experiments. The abscissae show the angular momentum given to the
halo, expressed as the usual dimensionless spin parameter.  Open
circles mark results from experiments in which the density profile of
the inner cusp was flattened, while squares indicate experiments where
cusp flattening did not occur.  Filled symbols show results from
experiments in which the MoI of the bar was increased by a factor 5 in
all cases except the point at the upper right, where the MoI was
increased 10-fold.  The changes to \Dv2\ make no allowance for halo
compression.  The bar parameters in each case are listed in
Table~\ref{ABWval}.}
\label{ABW}
\end{figure}

\section{Mean Density Reduction}
\label{reduction}
\subsection{Changes to \Dv2}
This study was motivated by the discrepancy, illustrated in
Fig.~\ref{Weiner}, between the predictions of halo density from LCDM
and that observed in real galaxies.  The solid and dashed lines in
that Figure show the predicted value of \Dv2\ (Alam \etal\ 2002) for
dark matter halos, that are generally above the observed points.  If
bar-halo friction could effect a reduction of the mean inner halo
density by about one order of magnitude, as measured by \Dv2, the
predicted lines could be shifted down by that factor and the
discrepancy between the predictions and the data would be largely
removed.

Since the halo parameters of mass and linear size in my simulations
can be scaled as desired, the only quantity of relevance that can be
extracted from them is the fractional change in \Dv2: $\mu =
\Dv2(t)/\Dv2(0)$.  Figure~\ref{ABW} shows the fractional change to the
inner halo density, $\mu$, measured from the simulations listed in
Table~\ref{ABWval}.  Results presented are exclusively from cases that
are numerically converged -- \ie\ results from low-$N$ simulations in
convergence tests are excluded.  Circles mark results from experiments
in which the density profile of the inner cusp was flattened.  Weaker
bars of any length lead to mild density reductions, as shown by the
points marked with squares.  The largest reductions to \Dv2, by a
factor $\mu^{-1} \ga 10$, occur when the inner part of the cusp is
flattened by exchanges with a long ($a=r_s$) bar.  Strong short bars
also flatten the cusp, but over a smaller volume, leading to a smaller
reduction in \Dv2.

The density reductions possible with rigid bars may underestimate the
largest that can be achieved, since real stellar bars are not rigid
objects with pattern speeds that decrease as dictated by a fixed
moment of inertia (MoI) as angular momentum is removed.  The stars
within the bar must lose angular momentum, but the pattern speed of
the bar is determined by the mean precession rate of the orbits.  (It
could even rise as the orbits shrink in size, although such behavior
has not been reported in any simulation, as far as I am aware.)  Thus
adopting the fixed MoI of a homogeneous ellipsoid may seriously
underestimate the angular momentum that could be extracted from the
bar.

Accordingly, I experimented with bars in which the effective MoI was
increased by a factor of five or ten from the standard value employed
so far, as noted in Table~\ref{ABWval}.  This stratagem resulted in a
correspondingly greater transfer of angular momentum to the halo over
a more protracted period as the pattern speed declined more slowly,
and the results are shown by the filled symbols in Figure~\ref{ABW}.
The enhanced MoI caused a greater reduction in the inner halo density
than in comparable experiments with the standard MoI, but by a
significant factor only if cusp flattening occurred.

A decrease in \Dv2\ by a factor $\ga 10$ requires a large ($a=r_s$),
massive, skinny bar, and the greatest changes occur when the MoI of
such bars are increased.  The density reduction by a shorter bar,
$a=0.2r_s$, is to about 60\% of the original \Dv2, which can be
boosted to $\sim 45\%$ by increasing the MoI.
\Ignore{ The modest changes to the rotation curve for these short bar
cases are shown in Figure~\ref{vrot}.}

\subsection{Angular Momentum Extracted from the Bar}
It is useful to express the angular momentum transferred to the halo
in terms of the usual dimensionless spin parameter, $\lambda =
LE^{1/2}/GM^{5/2}$.\Ignore{$=L_*/sqrt{12}$ for the Hernquist halo and
$=L_*/9.06$ for NFW, if $c=15$.}  Tidal torques lead to halos with a
log-normal distribution of spin parameters with a mean $\lambda \sim
0.05$.  Assuming, as usual, that the baryons and dark matter are well
mixed initially, the fraction of angular momentum in the baryons is
equal to the baryonic mass fraction: some 10\% -- 20\%.

The abscissae in Fig.~\ref{ABW} show the angular momentum transferred
to the halo, expressed as a change to $\lambda$.  Thus the angular
momentum that must be transferred from the baryons to the dark halo to
increase its spin parameter by $\Delta\lambda_{\rm b} \sim 0.01$
requires the removal of {\it all\/} the angular momentum that could
reasonably be expected to be possessed by the baryons!  This
conclusion suggests that no greater density reductions could be
achieved by this method.  Note that as the estimates of halo density
in Fig.~\ref{Weiner} are all from rotationally supported disks, these
galaxy disks must retain a significant fraction of their initial
angular momentum.

Since I have excluded the monopole term of the bar potential, and kept
the bar quadrupole fixed, these experiments ignore effects that
increase the halo density.  The halo must be compressed as baryons
cool and settle to make the disk, and contraction of a self-consistent
bar as it loses angular momentum can cause the halo to compress
further (Sellwood 2003; Col\'\i n \etal\ 2006), overwhelming any
density reduction caused by the angular momentum transferred.  Thus
the changes to \Dv2\ reported in Fig.~\ref{ABW} are likely to
be overestimates.

\section{Conclusions}
I have shown that reliable results can be obtained from careful
simulations of self-gravitating halos perturbed by a rigid bar without
the need for immense numbers of particles.  Rigid bars are an
idealization, but simplify the dynamics down to the bare essentials
over which disagreements remain.

Weinberg \& Katz (2007a) estimate the required numbers of particles
from perturbation theory.  Their ``coverage'' criterion is based on a
requirement that there be enough particles in a narrow range of
frequencies around the resonance to yield the correct statistical
balance between gainers and losers in resonant interactions.  Their
criterion, however, takes no account of the time-dependence of the
perturbation, which causes resonances to be broadened over a wide
range of frequencies allowing the correct response to be captured with
a much smaller $N$.  The excessive requirements suggested by WK07a
apply to the numerically much more delicate case of a steadily
rotating, fixed amplitude perturbation.  Furthermore, their diagnostic
diagrams require detailed balance at each point in $(E,L)$-space,
whereas the balance must be right for the complete ensemble of
resonant particles, which is a much larger fraction of the total.  My
Fig.~\ref{Lres} shows that simulations do indeed manifest resonant
exchanges with the perturbation that include a significant fraction of
the particles and the resonant response converges at moderate $N$.
Larger particle numbers enable the changes at resonances to be
illustrated in more detail, but the physical outcome of the
experiments is no different.

The minimum number of particles needed to obtain a converged result
does depend slightly upon the bar properties, and can be lowered by
adopting unequal mass particles.  I have shown that neither the
angular momentum transferred nor the halo density change varies as $N$
is increased above $\sim 10^5$ equal mass particles for long, massive,
skinny bars.  Simulations with over $10^8$ particles, which meet the
criteria suggested by WK07a, do not behave any differently from those
with three orders of magnitude fewer particles.  Shorter bars do
require more care than do large bars, but again I find the behavior
converges at moderate $N$, and that $10^8$ unequal mass particles is
far more than is needed.

Above this modest minimum number of particles, I find that results
from a grid code are identical to those obtained using the field
method devised by Hernquist \& Ostriker (1992), as explained in
\S\ref{discmeth}.  Results with different $N$, or with different
random seeds, show none of the stochasticity expected if there were
too few particles in any of the dynamically important resonances.

Mild bars, for which evolution is slower, require greater care; \eg,
my convergence test for the pattern speed evolution with self-gravity
(Fig.~13 of Paper I) indicated that $N \ga 10^6$ was required for a
very mild bar ($a=r_s$, $a:b = 1:0.5$, and $\Mb = 0.02$ or 8\% of the
enclosed halo mass).  However, such more delicate cases are incapable
of effecting a substantial density reduction (\S\ref{reduction}).

WK07a also invoke orbit scattering by density fluctuations as a second
reason to require large $N$.  In simulations where the halo density
reduction is substantial, the bar is slowed on an orbit timescale,
which is always much shorter than the relaxation timescale (Binney \&
Tremaine 1987), leading to a much lower particle number requirements
(\S\ref{noise}).  Furthermore, the experimentally determined mass
profiles are very smooth, and the radial acceleration will have
correspondingly little noise.  While this argument ignores
fluctuations in non-axisymmetric forces, I find my results are also
insensitive to changes in the order of azimuthal expansion
(Fig.~\ref{lmax}).

These results indicate that the estimates of the required numbers of
particles given by Weinberg \& Katz are greatly exaggerated.  The
evidence I have presented continues to indicate that careful
simulations with ${\cal O}(10^6)$ halo particles yield reliable
indications of the evolution of both the pattern speed and density
profile.

I have also determined the amount of halo density reduction that can
be brought about through angular momentum transfer from a strong,
initially rapidly-rotating bar, again with the limitation that the
simulations are not fully self-consistent.  My simulations with
massive, skinny bars confirm earlier work that the densities of cusped
DM halos can be reduced by bar-halo interactions.  However, I also
show that more moderate bars are able to achieve no more than a minor
reduction in the mean density of the inner halo, when halo compression
is neglected.  

I have found that large density reductions occur only when the inner
cusp is flattened to create a uniform density core, which I show
extends to a radius of about 1/3 the bar semi-major axis.  I have
demonstrated that flattening of the inner cusp is a collective
response of the halo that is driven by the bar torque.

In sequences of experiments in which the mass or axis ratio of the
rigid bar is gradually weakened, I find an abrupt change of behavior
from cusp flattening, to mild density reductions.  The sharp
transition as the bar quadrupole field is weakened appears to be real.
Behavior on either side of the sharp transition is independent of the
numerical parameters or the code used.  Pairs of simulations
straddling the boundary behave bimodally and results are never
intermediate.  Since I have found that triggering of the collective
effect in truly marginal cases does depend on numerical parameters,
the parameter values at which the outcome changes cannot be determined
precisely from simulations.  However, I emphasize again that the
outcome of all simulations reported in this paper is independent of
all numerical parameters, aside from an extremely narrow range around
this boundary.

A reduction of the mean inner density by an order of magnitude
requires a bar, having a semi-major axis equal to the break radius of
the halo density profile, \ie\ $\sim 12 - 20\;$kpc, axis ratio $a/b
\ga 3$, and {\it bar\/} mass $\ga 30\%$ of the enclosed halo mass.
Large reductions must be offset in part, and mild reductions
overwhelmed, by halo compression through baryonic settling, which has
not been included here.

Real bars probably have higher effective moments of inertia allowing
more angular momentum to be extracted from them.  Experiments to mimic
this effect resulted in somewhat larger density reductions for a given
bar; for reasonable bars, the overall density reduction remained less
than a factor two.  Extreme bars with enhanced moments of inertia also
achieved greater density reductions, but at the cost of transfering
more angular momentum to the halo than the baryons are likely to
possess.

The angular momentum available in the baryons limits the density
reduction achievable by bars.  Since the galaxies for which halo
density measurements are available in Fig.~\ref{Weiner} are all still
rotationally supported, the baryons cannot have invested all their
angular momentum into halo density reduction.  External perturbers,
such as massive companions, undoubtedly contain more angular momentum
and energy in orbital motion, and therefore may seem to have the
potential to achieve greater reductions.  It should be noted, however,
that merging is a process already taken into account in the predicted
profiles, since individual halos generally result from a series of
mergers (\eg\ Wechsler \etal\ 2002).

The density reductions reported here are overestimates of those
possible in reality, since I did not include the monopole terms of the
bar field.  A massive disk, in which the bar forms, must have
compressed the halo as the baryons settle towards its center, and the
mean density of the inner halo will have risen by perhaps a factor of
three (\eg\ Sellwood \& McGaugh 2005).  Furthermore, loss of angular
momentum from the bar causes it to contract further, producing yet
more halo compression that may even overwhelm any reduction in halo
density resulting from the angular momentum transfer (Sellwood 2003;
Col\'\i n \etal\ 2006).

\acknowledgments
I thank Victor Debattista for a number of helpful comments and
suggestions throughout this project, and Doug Hamilton for a useful
conversation.  I also thank Stacy McGaugh, Juntai Shen, Kristine
Spekkens, and Ben Weiner for helpful comments on the manuscript.  This
work was supported by grants AST-0507323 from the NSF and NNG05GC29G
from NASA.

\appendix
\medskip
For a homogeneous bar, Weinberg (1985) adopts the approximate
quadrupole potential (his eq.~28)
\begin{equation}
\Phi_{\rm b}(r,\theta,\phi) = {b_1 r^2 \over 1 + (r/b_5)^5}
Y_{22}(\theta, \phi - \phi_0)
\label{wtquad}
\end{equation}
where $Y_{22}=\sqrt{15/(32\pi)}\sin^2\theta e^{2i(\phi-\phi_0)}$ and
$\phi_0$ is the phase of the bar.  For a bar with axes $a_1:a_2:a_3$
and density $\rho$, he chooses
\begin{equation}
b_1 = \pi G\rho \sqrt{8\pi \over 15}(A_1-A_2),
\label{b1}
\end{equation}
and
\begin{equation}
b_5^5 = a_1a_2a_3{2\over5}{a_2^2-a_1^2 \over A_1-A_2}.
\end{equation}
The dimensionless elliptic integrals, $A_i$ are defined by
Chandrasekhar (1969 Ch.~3, eq.~18):
\begin{equation}
A_i = a_1a_2a_3 \int_0^\infty {du \over (a_i^2 +u ) \Delta},
\end{equation}
with
\begin{equation}
\Delta^2 = (a_1^2 +u)(a_2^2+u)(a_3^2+u).
\end{equation}
Note that the expression for $b_1$, eq.~(\ref{b1}), is twice that
given in eq.~(46) of Weinberg (1985), in order to obtain the correct
variation in $\Phi_{\rm b}$ between the major and minor axes at small
$r$.

\begin{figure}[t]
\includegraphics[width=\hsize]{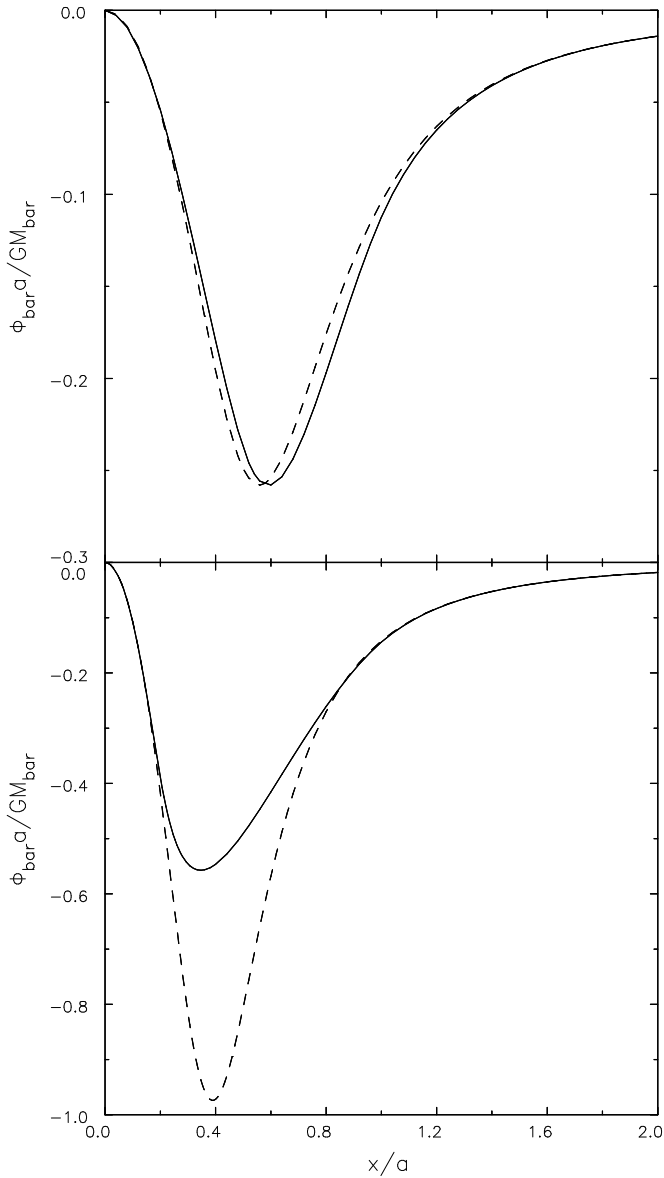}
\caption{\footnotesize The quadrupole part of the gravitational potential
along the major axis of a homogeneous bar with $a/b = 2$ (above) and
$a/b=5$ (below).  The solid curve gives the exact potential, the dashed
curve the approximation eq.~(\ref{quadrup}).  The approximation matches
well at small and large distances, but strongly overestimates the peak
for skinny bars.  Note the difference in scale of the ordinates between
the two panels.}
\label{quadapprox}
\end{figure}

I prefer to write eq.~(\ref{wtquad}) in the form (\cf\ eq.~\ref{quadrup})
\begin{equation}
\Phi_{\rm b}(r,\theta,\phi) = -{G\Mb \over a} {\alpha_2 r_*^2
\over 1 + (r_*/\beta_2)^5}\sin^2\theta e^{2i(\phi-\phi_0)},
\end{equation}
with $r_* = r/a$ and $a_1:a_2:a_3 = a : b : c$.  Comparing eqs.~(\ref{quadrup})
\& (\ref{wtquad}), we find that $\beta_2 = b_5/a$ and
\begin{equation}
\alpha_2 = {3a^2 \over 8bc} (A_1-A_2),
\end{equation}
since $\Mb = 4\pi abc\rho/3$.  Table~\ref{alpbet} gives the
values of $\alpha_2$ and $\beta_2$ for the bar axis ratios used in
this paper (\nb\ $a/c=10$ in all cases).

For completeness, the quadrupole potential in Cartesian coordinates
is:
\begin{equation}
\Phi_{\rm b}(x,y,z) = -{\alpha_2G\Mb \over a^3} {(x^2-y^2)\cos
      2\phi_0 + 2xy\sin 2\phi_0 \over 1 + (r/\beta_2a)^5}.
\end{equation}
Writing $\eta = r/(\beta_2a)$, $\nu = 1 + \eta^5$ and $\xi =
[(x^2-y^2)\cos2\phi_0 + 2xy\sin2\phi_0]/a^2$, this simplifies to
\begin{equation}
\Phi_{\rm b}(x,y,z) \simeq -{G\Mb \alpha_2 \over a} 
  {\xi \over \nu}.
\end{equation}
The acceleration components are
\begin{equation} 
a_x = {G\Mb \alpha_2 \over a^3} {2\nu(x\cos 2\phi_0 +
y\sin 2\phi_0) - 5\xi ax \eta^4/(r\beta_2) \over \nu^2} \\
\end{equation}
\begin{equation} 
a_y = {G\Mb \alpha_2 \over a^3} {2\nu(x\sin 2\phi_0 -
y\cos 2\phi_0) - 5\xi ay \eta^4/(r\beta_2) \over \nu^2} \\
\end{equation}
\begin{equation} 
a_z = -{G\Mb \alpha_2 \over a^3} {5\xi az \eta^4 \over
r\beta_2\nu^2}.
\end{equation}

Figure~\ref{quadapprox} compares the exact quadrupole potentials of
two homogeneous ellipsoids of different axis ratios with the
approximation given by eq.~(\ref{quadrup}); $a/c = 10$ in both cases.
The values of the parameters $\alpha_2$ and $\beta_2$ are defined to
ensure a good match at small and large distances for bars of any axis
ratio, which indeed they achieve.  While the approximation is pretty
good everywhere for the 2:1 bar (top panel), it increasingly
overestimates the peak strength of the quadrupole field as the bar
ellipticity increases, as shown for a 5:1 bar (bottom panel).

The exact field, which I used in Paper I, can be determined only
numerically, and therefore would not be easy for others to reproduce.
Throughout this paper, I have continued to use the approximation given
by eq.~(\ref{quadrup}), even though it clearly provides a stronger
perturbation than the nominal homogeneous bar when $a/b \gg 2$.  The
results continue to be of interest, however, since some other density
distribution could give rise to this stronger quadrupole.

It is unclear what form of the quadrupole WK07b adopted.  The text of
their paper states that they used the quadrupole approximation of
eq.~(\ref{quadrup}), which is the reason I adopted this expression,
but their Figure~3 shows the radial dependence for different axis
ratios on logarithmic scales.  Since the free parameters simply set
the amplitude and radius scales of the function, these curves all
ought to be self-similar, but they are not.  WK07b give no explanation,
but the deviations from the simple fitting function are in the correct
sense to provide a better fit to the exact field of a homogeneous
ellipsoid.

I have made repeated attempts to reproduce results from WK07b, using
NFW halos, including a rigid monopole term of the bar, and
experimenting with different approximations to the quadrupole, but
have not succeeded in reproducing the pattern speed or density
evolution they report for any of their simulations with skinny bars;
this contrasts with the success I had (Sellwood 2003) in reproducing a
result from Hernquist \& Weinberg (1992) for a rounder bar.  It seems
likely that the quadrupole field they used for the 5:1 bar in their
fiducial and other experiments has the form shown in their graph, and
not the functional form stated in their paper.

\end{document}